\documentclass[12pt]{article}
\usepackage{amsmath,epsfig,graphicx,verbatim,subfigure} 
\include{epsf} 
\def\ltap{\raisebox{-.6ex}{\rlap{$\,\sim\,$}} \raisebox{.4ex}{$\,<\,$}} 
\def\gtap{\raisebox{-.6ex}{\rlap{$\,\sim\,$}} \raisebox{.4ex}{$\,>\,$}}

\newcommand\as{\alpha_S}
\newcommand\f[2]{\frac{#1}{#2}} 
\def\beq{\begin{equation}} 
\def\eeq{\end{equation}} 
\def\beeq{\begin{eqnarray}} 
\def\eeeq{\end{eqnarray}} 
 
\def\to{\rightarrow}
 
\def\nn{\nonumber} 
 
\def\tL{{\widetilde L}}
 
\def\qt{q_T}
\def\mathbi#1{\textbf{\em #1}}
\def\bomega{\mathbf{\Omega}}
\def\bqt{\mathbi{q}\hspace*{-5pt}\phantom{q}_{\!{\mathbi T}}}
\def\bpt{\mathbi{p}\hspace*{-5pt}\phantom{p}_{\!{\mathbi T}}}
\def\bptj{\mathbi{p}\hspace*{-5pt}{\phantom{p}_{\!{\mathbi T}}}_j}
\def\bqtt{\mathbi{q}\hspace*{-5pt}\phantom{q}_{{\mathbi T}}}
\def\bkonet{\mathbi{k}\hspace*{-5pt}\phantom{k}_{{\!1\!\mathbi{T}}}}
\def\bktwot{\mathbi{k}\hspace*{-5pt}\phantom{k}_{{\!2\!\mathbi{T}}}}
\def\bkit{\mathbi{k}\hspace*{-5pt}\phantom{k}_{{i\!\mathbi{T}}}}

\begin{document} 
\begin{titlepage}
\vspace*{-20mm}
\begin{flushright}
ICAS 02/15\\
TIF-UNIMI-2015-10\\
ZU-TH 24/15
\end{flushright}
\renewcommand{\thefootnote}{\fnsymbol{footnote}}
\vspace*{3mm}

{\Large
\begin{center}
{\bf 
{Vector boson production at hadron colliders:\\[0.15cm]
transverse-momentum resummation\\[0.15cm] 
and leptonic decay}}
\end{center}}

\par \vspace{2mm}
\begin{center}
{\bf Stefano Catani${}^{(a)}$}, {\bf Daniel de Florian${}^{(b)}$},\\ 
\vskip .2cm
{\bf Giancarlo Ferrera${}^{(c)}$}
and
{\bf Massimiliano Grazzini${}^{(d)}$\footnote{On leave of absence from INFN, Sezione di Firenze, Sesto Fiorentino, Florence, Italy.}}\\
\vspace{5mm}
${}^{(a)}$INFN, Sezione di Firenze and
Dipartimento di Fisica e Astronomia,\\ Universit\`a di Firenze,
I-50019 Sesto Fiorentino, Florence, Italy\\
${}^{(b)}$Departamento de F\'\i sica, FCEYN, Universidad de Buenos Aires,\\
(1428) Pabell\'on 1 Ciudad Universitaria, Capital Federal, Argentina\\
and International Center for Advanced
Studies (ICAS), \\UNSAM, 
Campus Miguelete,  25 de Mayo y Francia,\\ 1650 Buenos
Aires, Argentina\\
${}^{(c)}$Dipartimento di Fisica, Universit\`a di Milano and\\ INFN, Sezione di Milano,
I-20133 Milan, Italy\\
${}^{(d)}$Physik-Institut, Universit\"at Z\"urich, CH-8057 Z\"urich, Switzerland
\end{center}

\begin{center} {\large \bf Abstract} \end{center}
\begin{quote}
\pretolerance 10000

We consider the transverse-momentum ($q_T$) distribution
of Drell--Yan lepton pairs produced, via $W$ and $Z/\gamma^*$ decay, in hadronic
collisions.
At small values of  $q_T$,
we resum the logarithmically-enhanced perturbative
QCD contributions
up to next-to-next-to-leading logarithmic accuracy. 
Resummed results are  consistently combined with the 
known ${\cal O}(\as^2)$ fixed-order results at intermediate and large values
of $q_T$.
Our calculation includes the leptonic decay of the vector boson 
with the corresponding spin correlations, the finite-width effects and
the full dependence on the final-state lepton(s) kinematics.
The computation is encoded in the numerical program {\tt DYRes}, which
allows the user to apply arbitrary kinematical cuts on the final-state leptons  
and to compute the corresponding distributions in the form of bin histograms.
We present a comparison of our results with some of
the available LHC data.
The inclusion of the leptonic decay in the resummed calculation
requires a theoretical discussion on the $q_T$ recoil due to the 
transverse momentum of the produced
vector boson. We present a $q_T$ recoil procedure 
that is directly applicable to 
$q_T$ resummed calculations for generic production processes of high-mass 
systems in hadron collisions.

\end{quote}

\vspace*{\fill}
\begin{flushleft}
July 2015

\end{flushleft}
\end{titlepage}

\setcounter{footnote}{1}
\renewcommand{\thefootnote}{\fnsymbol{footnote}}

\section{Introduction}
\label{sec:intro}

The production of high-mass lepton pairs through
the Drell--Yan (DY) mechanism \cite{Drell:1970wh} is a 
benchmark
hard-scattering process at hadron colliders.
It provides 
important tests of the Standard Model (SM) with 
both precise measurements of
its fundamental parameters and, at the same time, stringent constraints
on new physics.

It is thus a major task to achieve accurate theoretical predictions
for the DY production cross section and related kinematical distributions. 
This requires,
in particular, the evaluation of QCD radiative corrections,
which can be perturbatively computed as power series expansion in the 
strong coupling $\as$.
The total cross section \cite{Hamberg:1990np}
and the rapidity distribution \cite{Anastasiou:2003ds}
of the vector boson 
are known up to the next-to-next-to-leading order (NNLO)
in perturbative QCD.
Two independent fully 
exclusive
NNLO calculations, 
which include the leptonic decay of the vector boson, 
have been performed 
\cite{Melnikov:2006di,Catani:2009sm,Catani:2010en}.
Electroweak (EW) radiative corrections 
are also available
for both  $W$ \cite{ewW} and $Z/\gamma^*$ \cite{ewZ} production.
Mixed QCD-EW corrections have been considered in
Refs.~\cite{Kotikov:2007vr,Kilgore:2011pa,Bonciani:2011zz,Dittmaier:2014qza}.

A particularly relevant observable is the transverse-momentum ($q_T$) distribution of
the vector boson.
To obtain a precise measurement of the $W$ mass it is important to have
accurate theoretical calculations of the $W$ and $Z$ bosons $q_T$ spectra.
In the large-$q_T$ region ($q_T\sim m_V$), where the transverse momentum is
of the order of the vector boson mass $m_V$, QCD corrections are 
known up to ${\cal O}(\as^2)$
\cite{Ellis:1981hk,Arnold:1988dp,Gonsalves:1989ar}, and these results were extended in
Refs.\,\cite{Mirkes:1992hu,Mirkes:1994dp} with the inclusion of the dependence on the 
leptonic decay variables.
Very recently the fully exclusive 
${\cal O}(\as^3)$
computation of vector boson production 
in association with a jet has been performed in Ref.~\cite{Boughezal:2015dva}
(in the case of $W$ production) 
and Ref.~\cite{Ridder:2015dxa} (in the case of $Z/\gamma^*$ production).

%
The bulk of the vector boson cross section is produced in the small-$q_T$ region ($q_T\ll m_V$),
where the reliability of the fixed-order
expansion is spoiled by the presence of 
large logarithmic corrections,
$\as^n\,(m_V^2/q_T^2)\ln^m(m^2_V/q_T^2)$ (with $0\leq m \leq 2n-1$),
of soft and collinear origin.
To obtain reliable predictions, 
these logarithmically-enhanced terms 
have to be evaluated and systematically
resummed to all
orders in perturbation theory 
\cite{Dokshitzer:hw}--
\cite{Catani:2013tia}.
In recent years, the resummation of small-$q_T$ logarithms has been reformulated 
\cite{Gao:2005iu}--\cite{Echevarria:2015uaa}
by using Soft Collinear Effective Theory (SCET) methods 
and transverse-momentum dependent (TMD) factorization.

%
The resummed and fixed-order calculations, which are valid
at small and large values of  $q_T$, respectively, can be consistently 
matched at intermediate values of $q_T$
to achieve a uniform theoretical accuracy for the entire range of transverse momenta.

In this paper we compute the 
vector boson transverse-momentum distribution 
\cite{Bozzi:2008bb, Bozzi:2010xn} by using 
the resummation formalism proposed 
in Refs.~\cite{Catani:2000vq,Bozzi:2005wk,Bozzi:2007pn}, which 
can be applied to 
a generic process in which
a high-mass system of non strongly-interacting particles is produced 
in hadronic collisions
\cite{Bozzi:2003jy,Bozzi:2005wk,Bozzi:2007pn}, \cite{deFlorian:2011xf}--
\cite{Cieri:2015rqa}.
Other phenomenological studies of the DY 
$q_T$ distribution, which combine resummed and fixed-order perturbative
results at different levels of theoretical accuracy,
can be found in 
Refs.~\cite{Altarelli:1984pt}--\cite{D'Alesio:2014vja}.
Within the studies in Refs.~\cite{Altarelli:1984pt}--\cite{D'Alesio:2014vja},
the kinematical dependence on the momenta of the final-state leptons 
is considered
only in the {\tt RESBOS} calculation 
\cite{Balazs:1995nz,Balazs:1997xd,Guzzi:2013aja} and in the calculations of 
Refs.~\cite{Ellis:1997sc} and \cite{Banfi:2012du}.

Hadron collider experiments can directly measure only the decay products 
of vector bosons in finite kinematical regions.
Therefore,  it is important to include 
the vector boson leptonic decay in the theoretical calculations,
by retaining the kinematics of the final-state leptons.
In this way it is possible to obtain predictions for the transverse-momentum distribution 
of the measured leptons. This is 
specially relevant in the case of $W$ production where,  
because of the final-state neutrino, 
the transverse momentum of the vector boson can only be reconstructed through 
a measure of the hadronic recoil. 
Moreover, in both cases of $W$ and $Z$ production,
the inclusion of the leptonic decay allows one to apply kinematical 
selection cuts, thus
providing a more realistic simulation of the actual experimental analysis. 

In Ref.~\cite{Bozzi:2008bb,Bozzi:2010xn} we have presented a resummed 
computation of the transverse-momentum spectrum for $Z/\gamma^*$ production 
at Tevatron energies.
We have combined resummation at the next-to-next-to-leading logarithmic (NNLL)
accuracy in the small-$q_T$ region with the fixed-order results at 
${\cal O}(\as^2)$ in the large-$q_T$ region.
This leads to a calculation with uniform theoretical accuracy from small to
intermediate values of $q_T$. In particular, 
the integral over the range $0 \leq q_T \leq q_{T {\rm max}}$ 
($q_{T {\rm max}}$ is a generic upper limit in the small-$q_T$ region) 
of the $q_T$ distribution
includes the complete perturbative terms up to NNLO.
Moreover, 
at large values of $q_T$ the calculation  
implements a unitarity constraint that guarantees to exactly
reproduce the NNLO value of the total cross section after integration
over $q_T$.
In this paper we extend the NNLL+NNLO calculation of Ref.~\cite{Bozzi:2010xn}
to $W$ boson production, and we include the leptonic decay of the vector boson
with the corresponding spin correlations.
The spin of the vector boson dynamically correlates the decaying lepton momenta
with the transverse momentum acquired by the vector boson through its production
mechanism. Therefore, the inclusion of the full dependence 
on the lepton decay variables in the resummed calculation requires a theoretical
discussion on the treatment of the $q_T$ recoil due to the 
transverse momentum of the vector boson. We treat the $q_T$ recoil by 
introducing a general procedure that is directly applicable to 
$q_T$ resummed calculations for generic production processes of high-mass 
systems in hadron collisions.
The calculation presented in this paper parallels
the one performed in Ref.~\cite{deFlorian:2012mx} for the case
of SM Higgs boson production, with the non-trivial additional complication 
of dealing with the spin correlations that are absent in  
the Higgs boson case.
Our vector boson computation is implemented in the numerical code {\tt DYRes}, which allows the
user to apply arbitrary kinematical cuts on the final-state leptons and to compute the corresponding
relevant distributions in form of bin histograms. 
The code {\tt DYRes} is publicly available and it can be downloaded from the URL
address {\tt http://pcteserver.mi.infn.it/\char`\~ferrera/dyres.html}~.

The paper is organized as follows. In Sect.~\ref{sec:theory} we briefly review 
the resummation formalism of 
Refs.~\cite{Catani:2000vq,Bozzi:2005wk,Bozzi:2007pn},
and we discuss the main features of $q_T$ resummation for the DY process
with full dependence on final-state 
lepton variables. 
In Sect.~\ref{sec:results} we present our quantitative results for vector boson
production at LHC energies.
Section~\ref{sec:inclu} is devoted to the $q_T$ spectrum of the vector boson
after integration over the final-state leptons. We present results at different
orders of logarithmic accuracy, we study the corresponding dependence on scale
variations, and we briefly comment on uncertainties due to parton densities
and on non-perturbative effects.
In Sect.~\ref{sec:lepton}
we compare our numerical results
for $Z/\gamma^*$ and $W$ production 
with some of the available LHC data, and we also study 
the impact of transverse-momentum resummation on
lepton kinematical variables.
In Sect.~\ref{sec:summary} we summarize our results.
The Appendix
presents a detailed discussion of $q_T$ recoil and of its
implementation.

\section{Transverse-momentum resummation}
\label{sec:theory}

In this Section we briefly 
recall the main features of the transverse-momentum resummation formalism that 
we use in this paper. 
A more detailed discussion of the resummation formalism can be found in
Refs.~\cite{Catani:2000vq,Bozzi:2005wk,Bozzi:2007pn,Catani:2013tia}. 
In Ref.~\cite{Bozzi:2010xn} we have considered NNLL resummation for the
$q_T$ distribution of the vector boson after integration over the kinematical
variables of the decaying leptons and the rapidity of the vector boson.
In this paper we extend the results of Ref.~\cite{Bozzi:2010xn} to include the
entire kinematical dependence on the final-state leptons. The presentation in 
this Section parallels that of Sect.~2 in Ref.~\cite{Bozzi:2010xn} and, in
particular, we highlight the main differences that arise in the treatment
of the rapidity of the vector boson and, especially, of the lepton kinematics.

We consider the inclusive hard-scattering process
\begin{equation}
h_1(P_1) + h_2(P_2) \;\to\; V(q) 
+ X \;\to\; l_3(p_3)+ l_4(p_4) + X,   
\label{first}
\end{equation}
where the collision of the two hadrons $h_1$ and $h_2$ with momenta
$P_1^\mu$ and $P_2^\mu$ produces the vector boson $V$ ($V=W^+, W^-, Z$ and/or
$\gamma^*$)
with total momentum $q^\mu$,
which subsequently decays in the
lepton pair $l_3l_4$,
and $X$ denotes the accompanying final-state radiation. 
We consider high values of the invariant mass $M$ of the lepton pair
(in general, $M$ differs from the on-shell mass $m_V$ of the vector boson $V$),
and 
we treat the colliding hadrons and the leptons in the
massless approximation ($P_1^2=P_2^2=p_3^2=p_4^2=0$) throughout the paper.
In a reference frame where the colliding hadrons are back-to-back, 
the momentum $q^\mu$ is fully specified
by the invariant mass $M$ ($M^2=q^2$), 
the two-dimensional transverse-momentum vector $\bqt$ (with magnitude 
$q_T={\sqrt {\bqt^2}}$
and azimuthal angle $\phi_{\bqtt\,}$)
and the rapidity $y$ ($y=\frac12 \ln \frac{q\cdot P_2}{q\cdot P_1}$) 
of the vector boson.
Analogously the momentum $p_j^\mu$ of the lepton $l_j$ ($j=3,4$) is 
specified by the lepton rapidity
$y_j$ and transverse momentum $\bptj$.
 
The kinematics of the lepton pair is completely specified by six independent
variables (e.g., the three-momenta of the two leptons). 
For our purposes, it is convenient to use the vector boson momentum $q^\mu$ 
to select four independent variables. Therefore,
the final-state lepton kinematics is fully determined by the vector boson momentum $q^\mu=p_3^\mu+p_4^\mu$ and 
by two additional and {\em independent} variables
that specify the angular distribution of the leptons with respect to 
the vector boson momentum $q^\mu$.
We generically denote these two additional
kinematical variables as $\,\bomega=\{\Omega_A,\Omega_B\}$.
These two independent variables can be chosen in different ways.
For instance, we can use longitudinally boost invariant variables such as 
the rapidity difference $y_3-y$ and the azimuthal angle $\phi_3$ (or the
azimuthal angle difference
$\phi_3-\phi_{\bqtt\,}$)
of the lepton $l_3$ and the vector boson in 
the hadronic back-to-back reference frame.
Alternatively, we can use the polar and azimuthal angles $\{\theta',\phi'\}$ 
of one lepton
in a properly specified rest frame of the vector boson
(such as, for instance,
the Collins--Soper rest frame \cite{Collins:1977iv}).
Independently of the actual specification of the variables $\bomega$, the 
most general fully-differential hadronic cross 
section 
is expressed in terms of the sixfold differential distribution
\begin{equation}
\label{sixdim}
\frac{d\sigma_{h_1h_2\to l_3l_4}}{d^2{\bqt}\,d{M^2}\,dy\,d{\bomega}} 
(\bqt,M,y,s,{\bomega}) \;,
\end{equation}
where $s=(P_1+P_2)^2=2P_1 \cdot P_2$ is the square of the hadronic 
centre--of--mass energy. 
Obviously, the differential distribution also depends on the EW parameters
(including the mass $m_V$ of the vector boson $V$): unless otherwise specified,
this dependence is not explicitly denoted throughout the paper.

The differential hadronic cross 
section 
can be written as
\begin{eqnarray}
\frac{d\sigma_{h_1h_2\to l_3l_4}}{d^2{\bqt}\,d{M^2}\,dy\,d{\bomega}} (\bqt,M,y,s,{\bomega}) &=&
\sum_{a_1,a_2}\int_0^1dx_1\int_0^1dx_2 
\,f_{a_1/h_1}( x_1,\mu_F^2)\,f_{a_2/h_2}(x_2,\mu_F^2) \nonumber \\
\label{fact}
&\times&\,
{\frac{d{\hat\sigma}_{a_1a_2\to l_3l_4}}{d^2{\bqt}\,d{M^2}\,d\hat y\,d{\bomega}}} 
(\bqt,M,\hat y,\hat s,{\bomega};\alpha_S(\mu_R^2),\mu_R^2,\mu_F^2)\,,
\end{eqnarray}
where $f_{a/h}(x,\mu_F^2)$ ($a=q_f,\bar q_f,g$) are the parton densities of the colliding hadron $h$ 
at the factorization scale $\mu_F$, 
$d\hat\sigma_{a_1a_2\to l_3l_4}$ are the 
differential partonic cross sections, 
$\hat s = x_1 x_2 s$ is 
the square of the partonic centre--of--mass energy, 
$\hat y=y-\ln\sqrt{x_1/x_2}$ is the vector boson rapidity with respect to the
colliding partons,
and $\mu_R$ is the renormalization scale. 
Note that the partonic cross sections do not have any explicit dependence on
hadronic kinematical variables, since the leptonic variables ${\bomega}$ are
specified with respect to $q^\mu$.
The partonic cross section $d\hat\sigma_{a_1a_2\to l_3l_4}$ is computable in
QCD perturbation theory as a power series expansion in the QCD coupling $\as$.

In the region where $q_T\sim M$, the perturbative expansion of the partonic cross
section starts at ${\cal O}(\as)$. In this region
the value of the auxiliary scales $\mu_F$
and $\mu_R$ can be chosen to be of the order of $M$, and
the QCD perturbative
series is controlled by a small expansion parameter $\alpha_S(M^2)$.
Therefore, fixed-order calculations of the partonic cross section are
theoretically justified.
The QCD radiative corrections are known analytically 
up to ${\cal O}(\as^2)$ after integration over the lepton angular variables
\cite{Arnold:1988dp,Gonsalves:1989ar}
and with the inclusion of the full dependence on these angular variables
\cite{Mirkes:1992hu,Mirkes:1994dp}.  The numerical results at ${\cal O}(\as^2)$
can be obtained also from the fully-exclusive calculations of 
Refs.~\cite{Melnikov:2006di,Catani:2009sm,Catani:2010en}.
Results at ${\cal O}(\as^3)$ can be derived from the recent numerical 
computations
of $W+{\rm jet}$ production \cite{Boughezal:2015dva} and 
$Z/\gamma^*+{\rm jet}$ production \cite{Ridder:2015dxa}.

In the small $q_T$ region ($q_T\!\ll\! M$),
the perturbative computation of the partonic cross section starts at 
${\cal O}(\as^0)$ through the leading-order (LO) 
EW
process $q_f{\bar
q_{f^\prime}} \to V$ of quark--antiquark annihilation. In this region,
the QCD radiative corrections are known up to NNLO (i.e., ${\cal O}(\as^2)$)
in analytic form \cite{Catani:2012qa} 
by neglecting corrections of ${\cal O}(q_T/M)$
(these corrections can directly be extracted from 
Refs.~\cite{Arnold:1988dp,Gonsalves:1989ar,Mirkes:1992hu,Mirkes:1994dp}).
The complete (i.e., by including
corrections of ${\cal O}(q_T/M)$) NNLO result can be obtained from the numerical
computations of Refs.~\cite{Melnikov:2006di,Catani:2009sm}.
However, in the small $q_T$ region
the convergence of the fixed-order
perturbative expansion is spoiled
by the presence 
of powers of large logarithmic terms, 
$\as^n\; (M^2/q_T^2) \ln^m (M^2/q_T^2)$ (with $0\le m\le 2n-1$).
In particular, these terms become singular in the limit $q_T \to 0$.
To obtain reliable predictions these terms have to be resummed to all orders.

Within our formalism, the resummation is performed at the level of the partonic cross section, which is decomposed 
as follows:
\begin{equation}
\label{resplusfin}
\Big[d{\hat \sigma}_{a_1a_2\to l_3l_4}\Big]=
\Big[{d\hat \sigma}^{(\rm res.)}_{a_1a_2\to l_3l_4}\Big]
\,+\,\Big[d{\hat \sigma}^{(\rm fin.)}_{a_1a_2\to l_3l_4}\Big]\, .
\end{equation}
Here we have introduced a shorthand notation: the symbol 
$\Big[d{\hat \sigma}_{a_1a_2\to l_3l_4}\Big]$ denotes the multidifferential
partonic cross section that appears as the last factor in the right-hand side of
Eq.~(\ref{fact}).
The first term, ${d\hat \sigma}^{(\rm res.)}$,
on the right-hand side of Eq.~(\ref{resplusfin}) is the {\em resummed} component.
It contains all the logarithmically-enhanced contributions (at small $q_T$)
that have to be resummed
to all orders in $\alpha_S$.
The second 
term, the {\em finite} component $d{\hat \sigma}^{(\rm fin.)}$,
is free of such contributions and thus it
can be evaluated at fixed order in perturbation theory. 
Note that part of the non-singular (i.e., not logarithmically-enhanced)
contributions can also be included in ${d\hat \sigma}^{(\rm res.)}$, and 
we comment later about this point.

The resummation of the logarithmic contributions has to be carried out
in the impact parameter ($b$) space 
\cite{Dokshitzer:hw,Collins:1981uk,Collins:va,Altarelli:1984pt,Collins:1984kg}
to fulfil the important constraint of
transverse-momentum conservation for inclusive multiparton radiation.
The impact parameter $\bf b$ is the conjugate variable to $\bqt$ through a
Fourier transformation.
The small-$q_T$ region ($q_T \ll M$) corresponds to the large-$b$ region
($bM \gg 1$) and the logarithmic terms $\ln (M^2/q_T^2)$ become large 
logarithmic contributions $\ln (M^2b^2)$ in $b$ space. 
The resummed component of the cross section is then obtained by performing the 
inverse Fourier transformation (or the Bessel transformation in Eq.~(\ref{resw}))
from $\bf b$ space to $\bqt$ space.
The resummed component 
of the partonic cross section in Eq.~(\ref{resplusfin}) 
can be expressed as
\begin{equation}
\label{resum}
\left[ {d{\hat \sigma}_{a_1a_2\to l_3l_4}^{(\rm res.)}} \right]
= \sum_{b_1,b_2=q_f,{\bar q}_{f^\prime}} \!\!
\frac{{d{\hat \sigma}^{(0)}_{b_1b_2\to l_3l_4}}}{d{\bomega}} \;
\frac{1}{\hat s} \;
{\hat W}_{a_1a_2,b_1b_2\to V}(q_T^2,M,\hat y,\hat
s;\alpha_S(\mu_R^2),\mu_R^2,\mu_F^2) \;,
\end{equation}
where
\begin{equation}
\label{resw}
{\hat W}_{a_1a_2,b_1b_2\to V}(q_T^2,M,\hat y,\hat s;\alpha_S,\mu_R^2,\mu_F^2)
= 
\int_0^\infty \frac{db}{2\pi} \; b \,J_0(b q_T) 
\;{\cal W}_{a_1a_2,b_1b_2\to V}(b,M,\hat y,\hat s;\alpha_S,\mu_R^2,\mu_F^2) \;,
\end{equation}
and $J_0(x)$ is the $0$th-order Bessel function.
The factor $d{\hat \sigma}^{(0)}_{b_1b_2\to l_3l_4}$ in the right-hand side
of Eq.~(\ref{resum})
is the Born level differential cross section for the partonic subprocess
$q_f\bar q_{f'}\to V \to l_3l_4$ of 
quark--antiquark annihilation, where the quark flavours 
$f$ and $f'$  can be either different (if $V=W^\pm$) or equal 
(if $V=Z,\gamma^*$). This factor is of purely 
EW origin, 
and it completely encodes the dependence on
the lepton  kinematical variables 
${\bomega}$. We postpone more detailed comments on $d{\hat
\sigma}^{(0)}$ (see Eq.~(\ref{Frecoil}) and the discussion therein).
The QCD radiative corrections and their associated dependence
on $\ln (M^2b^2)$ are embodied in the 
resummed factor ${\cal W}_{a_1a_2,b_1b_2\to V}$,
which depends on the produced vector boson $V$ but it is independent
of the decay leptons (in particular, it does not depend on ${\bomega}$).
The integrand ${\cal W}$ in Eq.~(\ref{resw}) depends on ${\bf b}^2=b^2$ and the
inverse Fourier transformation is recast in terms of the Bessel transformation 
through the integration over the azimuthal angle of $\bf b$.
Note that the resummation factor ${\hat W}_{a_1a_2,b_1b_2\to V}$
depends on  $\qt^2$ and it does not contain any dependence on the 
azimuthal angle $\phi_{\bqtt\,}$ of $\bqtt\,$. This azimuthal independence is a
feature of transverse-momentum resummation \cite{Collins:1984kg} 
for the production processes
of colourless systems (such as vector bosons) through quark--antiquark
annihilation. In contrast, logarithmically-enhanced azimuthal correlations 
enter transverse-momentum resummation for processes initiated by gluon-gluon 
fusion \cite{Catani:2010pd} (such as Higgs boson production) 
and for production of systems that carry
colour charges (such as heavy quarks) \cite{Catani:2014qha} 
through either quark--antiquark
annihilation or gluon-gluon fusion.

The all-order resummation structure of ${\cal W}_{a_1a_2,b_1b_2\to V}$ 
in Eq.~(\ref{resw}) can
be organized in exponential form \cite{Bozzi:2005wk,Bozzi:2007pn}.
The exponentiated structure is directly evident by considering 
the `double' $(N_1,N_2)$ Mellin moments ${\cal W}_{V}^{(N_1,N_2)}(b,M)$ 
of the function
${\cal W}_{V}(b,M,\hat y,\hat s)$
with respect to the variables 
$z_1=e^{+\hat y}M/{\sqrt{\hat s}}$ and $z_2=e^{-\hat y}M/{\sqrt{\hat s}}$ 
at fixed $M$. We have~\footnote{For the sake of simplicity, in this presentation
we omit the explicit dependence on the parton indices $\{ a_1a_2,b_1b_2 \}$.
This simplified notation applies to the case of a sole parton species or, more
precisely, to flavour non-singlet partonic channels
(see
Refs.~\cite{Bozzi:2005wk,Bozzi:2007pn} for the general case).}
\begin{eqnarray}
\label{wtilde}
{\cal W}_{V}^{(N_1,N_2)}(b,M;\alpha_S(\mu_R^2),\mu_R^2,\mu_F^2)
&=&{\cal H}_V^{(N_1,N_2)}\left(M; 
\alpha_S(\mu_R^2),M/\mu_R,M/\mu_F,M/Q \right) \nonumber \\
&\times& \exp\{{\cal G}^{(N_1,N_2)}(\alpha_S(\mu_R^2),\tL;M/\mu_R,M/Q)\}
\;,
\end{eqnarray}
where 
the dependence on $b$ (and on the large logarithm $\ln (M^2b^2)$)
is denoted by defining and introducing 
the logarithmic expansion parameter $\tL\equiv \ln ({Q^2 b^2}/{b_0^2}+1)$
with $b_0=2e^{-\gamma_E}$ ($\gamma_E=0.5772...$ 
is the Euler number).
The scale $Q\sim M$, 
named resummation scale \cite{Bozzi:2003jy}, 
which appears in the right-hand side of Eq.~(\ref{wtilde}), parametrizes the
arbitrariness in the resummation procedure.
Although ${\cal W}_{V}^{(N_1,N_2)}$ does not depend on $Q$ when
evaluated to all perturbative orders, its explicit dependence on $Q$
occurs when it is computed by truncation of the resummed
expression at some level of logarithmic accuracy (see Eq.~(\ref{exponent})). 
Variations of $Q$ around $M$ can thus be used to estimate the
size of yet uncalculated higher-order logarithmic
contributions.

The contribution $\exp\{{\cal G}^{(N_1,N_2)}\}$ 
in the right-hand side of Eq.~(\ref{wtilde}) includes the Sudakov form factor
and collinear-evolution terms. This contribution (which does not depend on 
the factorization scale $\mu_F$) is universal (i.e. process independent), 
namely, it is independent on the produced vector boson $V$ and, more generally,
it occurs in transverse-momentum resummation for all the processes that are
initiated by quark--antiquark annihilation at the LO level. 
The generalized form factor 
$\exp\{{\cal G}^{(N_1,N_2)}\}$ contains all
the terms that order-by-order in $\alpha_S$ are logarithmically divergent 
as $b \to \infty$ (or, equivalently, as $q_T\to 0$). 
The all-order expression of the form factor can be systematically expanded in
terms of functions $g^{(k)}(\as\tL)$ of the resummation parameter 
$\as(\mu_R^2)\tL$ (each function $g^{(k)}(\as\tL)$ resums terms $\as^n\tL^n$
and it is defined such that $g^{(k)}(0)=0$).
The resummed logarithmic expansion of ${\cal G}^{(N_1,N_2)}$ in powers of 
$\as(\mu_R^2)$ reads
\begin{eqnarray}
\label{exponent}
{\cal G}^{(N_1,N_2)}(\alpha_S(\mu_R^2),\tL;M/\mu_R,M/Q)&=&\tL\, g^{(1)}(\alpha_S(\mu_R^2) \tL)
+g^{(2)\,(N_1,N_2)}(\alpha_S(\mu_R^2) \tL;M/\mu_R,M/Q) \nn \\
&+&\frac{\alpha_S(\mu_R^2)}{\pi} \;g^{(3)\,(N_1,N_2)}(\alpha_S(\mu_R^2) \tL;M/\mu_R,M/Q)+\dots \;,
\end{eqnarray}
where the term $\tL\, g^{(1)}$ collects the leading logarithmic (LL) 
contributions, the function $g^{(2)}$ includes
the next-to-leading logarithmic
(NLL) contributions \cite{Kodaira:1981nh},
$g^{(3)}$ controls the NNLL 
terms
\cite{Davies:1984sp, Becher:2010tm}
and so forth. 
The function ${\cal H}_V^{(N_1,N_2)}$ depends on the specific process of vector
boson production and it is due to hard-virtual and collinear contributions.
This function does not 
depend on the impact parameter $b$ (it includes all the perturbative
contributions to ${\cal W}_{V}^{(N_1,N_2)}$
that behave as constants in the limit $b\to\infty$) and, therefore,  
it can be expanded in powers of $\alpha_S=\alpha_S(\mu_R^2)$ as
\begin{equation}
\label{hexpan}
{\cal H}_V^{(N_1,N_2)}(M;\alpha_S)=
1+ \frac{\alpha_S}{\pi} \,{\cal H}_V^{(1)\,(N_1,N_2)} 
+ \left(\frac{\alpha_S}{\pi}\right)^2 
\,{\cal H}_V^{(2)\,(N_1,N_2)}+\dots \;.
\end{equation}
The next-to-leading order (NLO) term  ${\cal H}_V^{(1)\,(N_1,N_2)}$
is known since a long time \cite{Davies:1984hs}, and the NNLO term
${\cal H}_V^{(2)\,(N_1,N_2)}$ has been obtained more recently
by two independent calculations in Refs.~\cite{Catani:2012qa} and 
\cite{Gehrmann:2012ze}.
The explicit form of the functions
${\cal G}^{(N_1,N_2)}$  and ${\cal H}_V^{(1)\,(N_1,N_2)}$ and, in particular,
their dependence on the Mellin moment indices $(N_1,N_2)$
can be found in Ref.~\cite{Bozzi:2005wk} 
and in Appendix A of Ref.~\cite{Bozzi:2007pn}.

Incidentally, we recall that the generalized form factor 
$\exp\{{\cal G}\}$ is known up to NNLL accuracy also for processes initiated by
the gluon fusion mechanism 
\cite{Catani:vd,deFlorian:2000pr,Catani:2010pd,Becher:2010tm},
and that the ${\cal O}(\as^2)$ collinear coefficients (which contribute to 
the NNLO term in Eq.~(\ref{hexpan})) are also known for all possible partonic
channels 
\cite{Catani:2011kr,Catani:2012qa,Gehrmann:2012ze,Gehrmann:2014yya,
Catani:2013tia}. 
Owing to the universality structure of transverse-momentum
resummation, these results and those for the $q{\bar q}$ annihilation channel
(which contribute to vector boson production) can be directly implemented in
resummed calculations for production processes of generic high-mass systems.

The finite component $d{\hat \sigma}^{({\rm fin.})}$ in Eq.~(\ref{resplusfin})
has to be evaluated
starting from the usual fixed-order perturbative truncation 
of the partonic cross section
and subtracting the expansion of the resummed part at the same 
perturbative order. We have
\begin{equation}
\label{fincomp}
\Big[d{\hat \sigma^{({\rm fin.})}_{a_1a_2\to l_3l_4}}\Big]_{\rm f.o.}=
\Big[d{\hat \sigma_{a_1a_2\to l_3l_4}}\Big]_{\rm f.o.}
\,-\,
\Big[d{\hat \sigma^{({\rm res.})}_{a_1a_2\to l_3l_4}}\Big]_{\rm f.o.}\,,
\end{equation}
where the subscript f.o.\ denotes the perturbative truncation at the order f.o.\
(NLO, NNLO and so forth). 
The customary fixed-order component $\left[d{\hat \sigma_{a_1a_2}}\right]_{\rm f.o.}$
(and consequently also the  finite component) definitely contains azimuthal
correlations with respect to $\bqtt\,$,
although these are not logarithmically-enhanced in the small-$q_T$ region.

To obtain NLL+NLO accuracy we have to include the functions $g^{(1)}$  and 
$g^{(2) (N_1,N_2)}$ in the generalized form factor 
${\cal G}^{(N_1,N_2)}$ of Eq.~(\ref{exponent}), the 
function ${\cal H}_V^{(1) (N_1,N_2)}$ 
in the hard/collinear factor ${\cal H}_V^{(N_1,N_2)}$ of Eq.~(\ref{hexpan})
and the finite component of Eq.~(\ref{fincomp}) up to
${\cal O}(\as)$. To reach NNLL+NNLO accuracy we
need to include also the functions $g^{(3) (N_1,N_2)}$, 
${\cal H}_V^{(2) (N_1,N_2)}$ and
the finite component up to ${\cal O}(\as^2)$\footnote{This classification of the
resummed+matched expansion exactly coincides with that of 
Refs.~\cite{Bozzi:2005wk,Bozzi:2010xn}. We simply note that we are using labels
that differ from those used in Refs.~\cite{Bozzi:2005wk,Bozzi:2010xn}.
The various terms of the expansion are denoted here 
(analogously to Ref.~\cite{deFlorian:2012mx}) with the labels 
NLL+NLO and NNLL+NNLO, whereas they were denoted in 
Refs.~\cite{Bozzi:2005wk,Bozzi:2010xn} with the corresponding 
labels NLL+LO and NNLL+NLO. The fixed-order labels NLO and NNLO used here
directly refer to the perturbative accuracy in the small-$q_T$ region
(which corresponds to the perturbative accuracy of the total cross section), 
whereas the labels LO and NLO used in
Refs.~\cite{Bozzi:2005wk,Bozzi:2010xn} were directly referring to 
the perturbative accuracy in the large-$q_T$ region.}.
This matching procedure 
between resummed and finite contributions guarantees
to achieve uniform theoretical accuracy 
over the entire range
of transverse momenta.
In particular, we remark that  
the inclusion of ${\cal H}_V^{(2) (N_1,N_2)}$ in the resummed
component at the NNLL+NNLO level is essential to achieve NNLO accuracy
in the small-$q_T$ region (considering a generic upper limit value
$q_{T {\rm max}}$,
the integral over the range 
$0 \leq q_T \leq q_{T {\rm max}}$ of the $q_T$ distribution
at the NNLL+NNLO level includes the complete perturbative terms up to NNLO).
An analogous remark applies to the inclusion of ${\cal H}_V^{(1) (N_1,N_2)}$
at the NLL+NLO level.

We have so far illustrated the resummation formalism for the most general
sixfold differential partonic cross section   
$\Big[d{\hat \sigma_{a_1a_2\to l_3l_4}}\Big]$
(and for the corresponding hadronic cross section in Eq.~(\ref{fact})).
Starting from $\Big[d{\hat \sigma_{a_1a_2\to l_3l_4}}\Big]$
and performing integrations over some kinematical variables,
we can obtain resummed results for more inclusive $q_T$-dependent distributions.
For instance, integrating over the lepton kinematical variables $\bomega$, 
we obtain the $q_T$ cross section  $d\sigma/(d^2{\bqt}\,d{M^2}\, dy)$ at
fixed invariant mass and rapidity of the lepton pair. The corresponding resummed
component of the partonic cross section, as obtained from Eq.~(\ref{resum}), is
\begin{equation}
\label{resumydiff}
{\frac{d{\hat\sigma}_{a_1a_2\to l_3l_4}^{({\rm res.)}}}{d^2{\bqt}\,d{M^2}\,d\hat y}} 
(\bqt,M,\hat y,\hat s;\alpha_S,\mu_R^2,\mu_F^2)
=\!\!\!\!\!\!\!\!\!\sum_{b_1,b_2=q_f,{\bar q}_{f^\prime}} \!\!\!\!\!
{\hat \sigma}^{(0)}_{b_1b_2\to l_3l_4}(M^2) \;  
\frac{1}{\hat s} \,
{\hat W}_{a_1a_2,b_1b_2\to V}(q_T^2,M,\hat y,\hat
s;\alpha_S,\mu_R^2,\mu_F^2) ,
\end{equation}
where ${\hat \sigma}^{(0)}_{q_f {\bar q}_{f^\prime} \to l_3l_4}(M^2)$
is the Born level 
(EW) total cross section for the partonic subprocess
$q_f\bar q_{f'}\to V \to l_3l_4$.
By performing an additional integration over the rapidity $y$ of the 
vector boson (lepton pair), we obtain $d\sigma/(d^2{\bqt}\,d{M^2})$
and the corresponding resummed component of the partonic cross section
simply involves the integration over $\hat y$ of the resummed factor
${\hat W}(q_T^2,M,\hat y,\hat s)$ in Eqs.~(\ref{resum}) and (\ref{resumydiff})
(or, equivalently, the factor 
${\cal W}(b,M,\hat y,\hat s)$ in Eq.~(\ref{resw})).
After integration over $\hat y$, the ensuing resummed factor depends on $M$ and 
$\hat s$, and it can be conveniently expressed in exponentiated form 
\cite{Bozzi:2005wk}
by considering `single' $N$ Mellin moments with respect to the variable
$z=M^2/\hat s$ at fixed $M$. The resummed expression for these 
`single' $N$ moments is exactly obtained by simply setting $N_1=N_2=N$ in
Eqs.~(\ref{wtilde})--(\ref{hexpan}).
Our resummed calculation of $d\sigma/(d^2{\bqt}\,d{M^2})$ was discussed in
Ref.~\cite{Bozzi:2008bb,Bozzi:2010xn}, and it is implemented in the numerical 
code {\tt DYqT}. In Refs.~\cite{Bozzi:2008bb,Bozzi:2010xn} we presented detailed
quantitative results for vector
boson production at Tevatron energies. Results from {\tt DYqT}
at LHC energies are presented in the following Sect.~\ref{sec:inclu}.

Within our formalism the resummation of the large terms $\ln(M^2/q_T^2)$
at small values of $q_T$ is achieved by first performing the Fourier
transformation of the $q_T$ cross section (or, more precisely, of its singular
behaviour in the small-$q_T$ region) from $q_T$ space to $b$ space (incidentally,
the renormalization scale $\mu_R$ and the others auxiliary scales $Q$ and $\mu_F$
are kept fixed and, especially, {\em independent } of $q_T$ in the integration over
$q_T$ of the Fourier transformation). In $b$ space, the large logarithmic variable
(whose dependence has to be resummed) is $\tL$, at large values of $b$.
Note that in the context of the resummation approach, the parameter 
$\as(\mu_R^2) \tL$ is formally considered to be of order unity. Therefore, the
ratio of two successive terms in the expansion (\ref{exponent}) is 
formally of ${\cal O}(\as(\mu_R^2))$ (with no $\tL$ enhancement). In this respect
the resummed logarithmic expansion in Eq.~(\ref{exponent}) is as systematic as any
customary fixed-order expansion in powers of $\as(\mu_R^2)$. Analogously to any
perturbative expansions, the perturbative terms 
$g^{(k)\,(N_1,N_2)}(\alpha_S(\mu_R^2) \tL;M/\mu_R,M/Q)$ in Eq.~(\ref{exponent})
have an explicit logarithmic dependence on $\ln(M/\mu_R)$ or $\ln(M/Q)$
(see, e.g., Eqs.~(22) and (23) in Ref.~\cite{Bozzi:2005wk}). Therefore,
to avoid additional large logarithmic enhancements that would spoil the formal
behaviour of the expansion in Eq.~(\ref{exponent}), the renormalization scale 
$\mu_R$ has to be set at a value of the order of $M \sim Q$. A completely
analogous reasoning applies to the $\mu_F$ dependence of 
${\cal H}_V^{(N_1,N_2)}
\left(M; \alpha_S(\mu_R^2),M/\mu_R,M/\mu_F,M/Q \right)$ in the expansion of
Eq.~(\ref{hexpan}) and, therefore, we should set $\mu_F \sim M$.
In other words, once the enhanced perturbative dependence on $b^2M^2$ (i.e., on
the two different scales $M$ and $1/b$) is explicitly resummed (albeit at a
definite logarithmic accuracy), we are effectively dealing with a single-scale
observable at the hard scale $M$ and we can set $\mu_R \sim \mu_F \sim M$ in both
the resummed and finite components of the $q_T$ cross section in 
Eq.~(\ref{resplusfin}).

We remark that setting $\mu = {\cal O}(M)$ (here $\mu$ generically denotes the
auxiliary scales $\mu_R, \mu_F, Q$) does not mean that the $q_T$ cross section is
physically controlled by parton radiation with intensity that is proportional to
$\as(M^2)$. The resummed form factor $\exp\{{\cal G}^{(N_1,N_2)}\}$ in 
Eq.~(\ref{wtilde}) (and the ensuing logarithmic expansion in Eq.~(\ref{exponent}))
is produced by multiparton radiation with intensity that is proportional to
$\as(k^2)$ and $k^2$ is a dynamical scale that varies in the range 
$M^2 > k^2 > 1/b^2$ (see, for instance, Eq.~(19) in Ref.~\cite{Bozzi:2005wk}),
where $1/b^2$ can be physically identified with $q_T^2$ at small values of $q_T$.
Setting $\mu \sim M$ in Eqs.~(\ref{wtilde}) and (\ref{exponent}) corresponds,
roughly speaking, to consider the scale range $\mu^2 > k^2 > 1/b^2$
(it does not correspond to set $k^2 \sim \mu^2 \sim M^2$).

We recall \cite{Bozzi:2005wk} a feature of our resummation formalism. The
small-$q_T$ singular contributions that are resummed in Eqs.~(\ref{resum})
(or Eq.~(\ref{resumydiff})) are controlled by the large logarithmic parameter
$\ln(M^2/q_T^2)$, which corresponds to $L=\ln(Q^2b^2/b_0^2)$ (with $Q \sim M$) in
$b$ space at $b \to \infty$. In our resummation formula (\ref{wtilde}),
we actually use the logarithmic parameter 
$\tL = \ln ({Q^2 b^2}/{b_0^2}+1)$ \cite{Bozzi:2003jy}.
The motivations to use the logarithmic parameter $\tL$ are detailed in 
Ref.~\cite{Bozzi:2005wk} (see, in particular, the Appendix~B and the comments that
accompany Eqs.~(16)-(18) and Eqs.~(74)-(75) in Ref.~\cite{Bozzi:2005wk}), and here
we simply limit ourselves to recalling some aspects.
In the relevant resummation region $bQ \gg 1$, we have 
$\tL = L +{\cal O}(1/(Q^2b^2))$ and, therefore, $\tL$ and $L$ are fully equivalent
to arbitrary logarithmic accuracy (in other words, the replacement 
$\tL \leftrightarrow L$ simply modifies the partition of small-$q_T$ non-singular
contributions between the two components in the right-hand side of 
Eq.~(\ref{resplusfin})). 
However, $L$ and $\tL$ have a very different behaviour as
$b \to 0$ (and, thus, they differently affect the $q_T$ cross section
in the large-$q_T$ region\footnote{The contribution of the integral in 
Eq.~(\ref{resw}) from the integration region where 
$b \ltap {\cal O}(1/M) \sim{\cal O}(1/Q)$ always gives (provided ${\cal W}(b,M)$
is integrable over such region) a non-singular contribution to the 
$q_T$ cross section in the small-$q_T$ region.}). 
When $bQ \ll 1$, we have
$L \gg 1$ and, therefore, the replacement $\tL \to L$ in Eq.~(\ref{wtilde})
would produce the resummation of large and unjustified perturbative contributions
in the large-$q_T$ region (strictly speaking, the replacement $\tL \to L$
leads to a $q_T$ cross section that is even not integrable over  
$q_T$ when $q_T \to \infty$: see, in particular, Eqs.~(131) and (132) of the
arXiv version of Ref.~\cite{Bozzi:2005wk} and related accompanying comments). 
In contrast, when $bQ \ll 1$
we have $\tL \to 0$ and ${\cal G}^{(N_1,N_2)} \to 0$. Therefore, the use of
$\tL$ reduces the impact of unjustified large contributions that can be introduced
in the 
small-$b$
region through the resummation procedure. Moreover, the
behaviour of the form factor $\exp\{ {\cal G}^{(N_1,N_2)} \}$ at $b=0$ is related
to the integral over $q_T$ of the $q_T$-dependent cross section and, since
we have $\exp\{ {\cal G}^{(N_1,N_2)} \}=1$ at $b=0$, our resummation formalism
fulfils a perturbative {\em unitarity constraint} \cite{Bozzi:2005wk}: after inclusion of the
finite component as in Eq.~(\ref{fincomp}), the integration over $q_T$ of our
resummed $q_T$ cross sections recovers the fixed-order predictions for the total
cross sections. Specifically, the integral over $\bqt$ of
$d\sigma/(d^2{\bqt}\,d{M^2} dy)$ and $d\sigma/(d^2{\bqt}\,d{M^2})$ at the
NNLL+NNLO (NLL+NLO) accuracy completely and exactly (i.e., with no additional 
higher-order
contributions) agrees with the rapidity distribution
$d\sigma/(d{M^2} dy)$ and the total cross section $d\sigma/d{M^2}$
at NNLO (NLO) accuracy, respectively.
In summary, the expressions (\ref{wtilde}) and (\ref{exponent}) in terms of the
logarithmic parameter $\tL$ correctly resum the large parametric dependence on
$\ln (bQ)$ at large values of $bQ$ and they introduce parametrically-small
perturbative contributions at intermediate or small values of $bQ$ 
(the coefficients of the perturbative corrections are proportional to 
powers of $\tL$ with $\tL \sim {\cal O}(1)$ if $bQ\sim {\cal O}(1)$ or
$\tL \ll 1$ if $bQ\ll 1$).
After having combined the resummed calculation at N$^k$LL accuracy with the
complete N$^k$LO calculation, as in 
Eqs.~(\ref{resplusfin}) and (\ref{fincomp}),
these parametrically-small corrections produce residual terms that start to
contribute at the N$^{k+1}$LO level. Therefore, the use of $\tL$ has the purpose
of reducing the impact of unjustified and large higher-order 
(i.e., beyond the N$^k$LO level) contributions that can be possibly introduced 
at intermediate and large values of $q_T$ through the resummation of the
logarithmic perturbative behaviour at small values of $q_T$. 
In particular, no residual higher-order contributions are introduced in the case
of the total (integrated over $q_T$) cross section (which is the most basic
quantity that is not affected by logarithmically-enhanced perturbative
corrections). 

We add some relevant comments about the dependence of the resummed cross section
on the kinematical variables  ${\bomega}$ that specify the angular distribution 
of the leptons with respect to the vector boson. By direct inspection of
Eqs.~(\ref{resum}) and (\ref{resumydiff}) we see that they involve exactly the
same resummation factor ${\hat W}$. The only difference between the right-hand
side of these equations arises form the Born level factors 
$d{\hat \sigma}^{(0)}/d{\bomega}$ and ${\hat \sigma}^{(0)}$, which are related as
follows through the integration over $\bomega$:
\begin{equation}
\label{Frecoil}
\frac{{d{\hat \sigma}^{(0)}_{q_f {\bar q}_{f^\prime}
\to l_3l_4}}}{d{\bomega}} = 
{\hat \sigma}^{(0)}_{q_f {\bar q}_{f^\prime} \to l_3l_4}(M^2)
\; F_{q_f {\bar q}_{f^\prime} \to l_3l_4}(\bqt/M;M^2, {\bomega}) \;\;,
\end{equation}
with the normalization condition
\begin{equation}
\label{Fnorm}
\int d{\bomega} \;F_{q_f {\bar q}_{f^\prime} \to l_3l_4}(\bqt/M;{\bomega}) =1 \;.
\end{equation}
Although both factors depend on 
EW parameters 
(EW couplings,
mass and width of the vector boson), they have a different dependence on the
relevant kinematical variables. The vector boson distribution 
$d\sigma^{({\rm res.})}/(d^2{\bqt}\,d{M^2}\, dy)$ (and, analogously, 
$d\sigma^{({\rm res.})}/(d^2{\bqt}\,d{M^2})$) involves the Born level
total cross section ${\hat \sigma}^{(0)}(M^2)$, which depends on $M^2$,
whereas the less inclusive leptonic distribution 
$d\sigma^{({\rm res.})}/(d^2{\bqt}\,d{M^2}\, dy \,d{\bomega})$ involves the 
Born level differential cross section $d{\hat \sigma}^{(0)}/d{\bomega}$ that
additionally depends on ${\bomega}$ and {\em also} on the transverse momentum 
$\bqt$ of the lepton pair (see the function $F$ in the right-hand side of
Eq.~(\ref{Frecoil})). To our knowledge the $q_T$ dependence of 
$d{\hat \sigma}^{(0)}/d{\bomega}$ has not received much attention in the previous
literature on transverse-momentum resummation and, therefore, we discuss this
issue with some details in Appendix~\ref{app}.
Physically, this $q_T$ dependence is a necessary consequence of 
transverse-momentum conservation and it arises as a $q_T$-recoil effect in
transverse-momentum resummation. At the LO in perturbation theory the lepton
angular distribution is determined by the Born level production and decay process
of the vector boson, which carries a vanishing transverse momentum.
Through the resummation procedure at fixed lepton momenta, higher-order
contributions due to {\em soft} and {\em collinear} multiparton radiation
{\em dynamically} produce a finite value of the transverse momentum
$q_T$ of the lepton pair, and this finite value of $q_T$ has to be distributed
between the two lepton momenta by affecting the lepton angular distribution.
This $q_T$-recoil effect on the Born level angular distribution is a non-singular
contribution to the $q_T$ cross section at small values of $q_T$ and, therefore,
it is not directly and unambiguously computable through 
transverse-momentum resummation. In other words, the Born level function
$F$ in Eq.~(\ref{Frecoil}) has the form
\begin{equation}
\label{Frecoilform}
F_{q_f {\bar q}_{f^\prime} \to l_3l_4}(\bqt/M;M^2, {\bomega}) =
F_{q_f {\bar q}_{f^\prime} \to l_3l_4}({\bf 0};M^2, {\bomega})
+ {\cal O}(\bqt/M) \;\;,
\end{equation}
where $F({\bf 0};M^2, {\bomega})$ is uniquely determined, whereas the small-$q_T$
corrections of ${\cal O}(\bqt/M)$ has to be properly specified.
In any physical computations of lepton observables (i.e., in any 
computations that avoid possible unphysical behaviour due to violation of momentum
conservation for the decay process $q = p_3(l_3) + p_4(l_4)$) through
transverse-momentum resummation, a consistent $q_T$-recoil prescription has to be
actually (either explicitly or implicitly) 
implemented\footnote{The dynamical treatment of the $q_T$ recoil is embedded
in the formulation of transverse-momentum $(k_T)$ factorization 
\cite{Catani:1990xk,Deak:2009xt}
of hard-scattering
processes at high energy (at small $x$).
}. 
Note that, after having combined the resummed and finite components as in 
Eqs.~(\ref{resplusfin}) and (\ref{fincomp}), the ${\cal O}(\bqt/M)$ recoil effects
lead to contributions that start at ${\cal O}(\as^3)$ (i.e., N$^3$LO) in the case
of our resummed calculation at NNLL+NNLO accuracy
(correspondingly, these contributions start at ${\cal O}(\as^2)$ in the case of
NLL+NLO accuracy). Obviously there are infinite ways of implementing the 
$q_T$-recoil 
effect, and in Appendix~\ref{app} 
we explicitly describe a very general 
and consistent
procedure\footnote{The $q_T$-recoil issue is not a specific issue of leptonic
decay in vector boson production. The issue is completely general
(see Appendix~\ref{app}), and it arises in any $q_T$ resummed calculation for the
production of a set of particles with measured momenta at fixed total transverse
momentum $q_T$ (e.g., diphoton, diboson, or heavy-quark pair production).
A noticeable exception (as discussed in Appendix~\ref{app}) 
is the production of a SM Higgs boson and its subsequent decay. In this case,
due to the spin-0 nature of the Higgs boson, the $q_T$ dependence of the
corresponding Born level function 
$F(\bqt/M;M^2, {\bomega})$ can be entirely determined by kinematics
\cite{deFlorian:2012mx}, without the
necessity of specifying $q_T$-recoil effects of dynamical origin.}.
Note that the $q_T$-recoil effect completely cancels after integration over the
leptonic variables ${\bomega}$ (see Eq.~(\ref{Fnorm})).

Our resummed calculation of the sixfold differential distribution in 
Eq.~(\ref{sixdim}) is implemented in the numerical code {\tt DYRes},
which allows the
user to apply arbitrary kinematical cuts on the momenta of the
final-state leptons and 
to compute the corresponding
relevant distributions in form of bin histograms.
We add some comments on the numerical implementation of our calculation.
In Eqs.~(\ref{wtilde})--(\ref{hexpan})
we have illustrated the structure of the resummed component in the double
$(N_1,N_2)$ Mellin space. Through the inverse Mellin transformation, this
structure can equivalently be expressed in terms of convolutions with respect
to longitudinal momentum fractions $x_1$ and $x_2$ (see Eq.~(\ref{fact})).
In the {\tt DYRes} code, the Mellin inversion is carried out numerically.
The results for the NNLO term ${\cal H}_V^{(2)}$ in Eq.~(\ref{hexpan})
are presented in Ref.~\cite{Catani:2012qa}
in analytic form directly in $(x_1,x_2)$ space. These results have to be
transformed in Mellin space. Then,
the Mellin inversion requires
the numerical evaluation of some basic $N$-moment functions that appear in the
expression of ${\cal H}_V^{(2) (N_1,N_2)}$: this evaluation has to be performed
for complex values of $N$, and we use the numerical results of 
Ref.~\cite{Blumlein:2000hw}. This implementation of the resummed component is
completely analogous to that of the {\tt DYqT} code  
\cite{Bozzi:2008bb,Bozzi:2010xn} and of other previous computations \cite{Bozzi:2007pn}.
Nonetheless, the efficient generation of `vector boson events' according 
to the multidifferential distribution
of Eq.~(\ref{resplusfin}) and the inclusion of the leptonic decay are 
technically non
trivial, and this requires substantial improvements
in the computational speed of the numerical code that evaluates the resummed 
component of the cross section. The fixed-order (NLO and NNLO) cross section in
Eq.~({\ref{fincomp}) 
and then the finite component of the cross section in Eq.~(\ref{resplusfin})
are evaluated through an appropriate modification of the {\tt DYNNLO} 
code~\cite{Catani:2009sm}: {\tt DYNNLO}
is particularly suitable to this purpose, since it is based on the
$q_T$ subtraction formalism \cite{Catani:2007vq}, which uses the
transverse-momentum resummation formalism to construct the subtraction
counterterms.

Using the resummation expansion parameter ${\widetilde L}$ in
Eq.~(\ref{wtilde}) and the matching procedure (which implements the 
perturbative unitarity
constraint on the total cross section)
with the complete fixed-order calculation, 
our resummation formalism \cite{Bozzi:2005wk} formally achieves
a uniform theoretical accuracy in the region of small and intermediate values of
$q_T$, and it avoids the introduction of large unjustified higher-order
contributions in the large-$q_T$ region. In the large-$q_T$ region, the results
of the resummed calculation are consistent with the customary fixed-order results
and, typically \cite{Bozzi:2005wk, Bozzi:2010xn}, show larger theoretical
uncertainties (e.g., larger dependence with respect to auxiliary-scale
variations) with respect to the corresponding fixed-order results. This feature
is not unexpected, since the theoretical knowledge (and the ensuing resummation)
of large logarithmic contributions at small $q_T$ cannot improve the theoretical
predictions at large values of $q_T$. In the large-$q_T$ region, where the
resummed calculation shows `unjustified' large uncertainties and ensuing loss
of predictivity with respect to the fixed-order calculation, the reliability of
the resummed calculation is superseded by that of the fixed-order calculation.
In this large-$q_T$ region, we can simply use the theoretical results of the 
fixed-order calculation. In the computation of quantities that directly and
explicitly depend on $q_T$ (e.g., the transverse-momentum spectrum of the vector
boson), it is relatively straightforward to identify and select `a posteriori'
the large-$q_T$  region where the resummed calculation is superseded by 
the fixed-order calculation.
In the present work, however, we are also interested in studying kinematical 
distributions of the vector boson decay products:
our goal is thus to generate the {\em full} kinematics of the vector boson 
and its (leptonic) decay, to apply the required acceptance cuts, and to compute
the relevant distributions of the lepton kinematical variables. 
In this framework, the actual results can become sensitive to the large-$q_T$
region in which the resummed calculation 
cannot improve the accuracy of the fixed-order calculation.
To reduce this sensitivity, in 
the {\tt DYRes}
implementation of the 
resummed calculation
we thus introduce a smooth
switching procedure at large value of $q_T$ by replacing
the resummed cross section in Eq.~(\ref{resplusfin}) as follows:
\begin{equation}
\label{switch}
\Big[d{\hat\sigma}_{a_1a_2\to l_3l_4}\Big]
\to w(q_T) 
\left(\Big[{d\hat \sigma}^{(\rm res.)}_{a_1a_2\to l_3l_4}\Big]
\,+\,\Big[d{\hat \sigma}^{(\rm fin.)}_{a_1a_2\to l_3l_4}\Big]\right)
+(1-w(q_T)) \Big[d{\hat\sigma}_{a_1a_2\to l_3l_4}\Big]_{\rm f.o.}\,,
\end{equation}
where the function $w(q_T)$ is defined as
\begin{equation}
\label{wpt}
w(q_T) =
\bigg \{
\begin{array}{cl}
1 & q_T \leq q_T^{\rm sw.}\\
f(q_T) & q_T > q_T^{\rm sw.}\\
\end{array}
\end{equation}
and the function $f(q_T)$ is chosen as
\begin{equation}
\label{fswitch}
f(q_T)=\exp\Big\{-\f{\left(q_T^{\rm sw.}-q_T\right)^2}{2(\Delta q_T)^2}\Big\}\, .
\end{equation}
We have quantitatively checked that the value of the parameter 
$q_T^{\rm sw.}$ can be
`suitably' chosen in the large-$q_T$ region, and that both 
parameters $q_T^{\rm sw.}$ and $\Delta q_T$ can be consistently chosen
so as not to spoil our unitarity constraint (in Sect.~\ref{sec:inclu}
we show that the integral over $q_T$ of our NLL+NLO and NNLL+NNLO
resummed results still reproduces well the NLO and NNLO total cross
sections). We note that we do not introduce any switching procedure in the
{\tt DYqT} calculation (though, its introduction is feasible) since, as
previously mentioned, the identification of the large-$q_T$ region is
straightforward in the computation of $d\sigma/(d^2\bqt \, dM^2)$. 

We recall \cite{Bozzi:2005wk} that the resummed form factor
$\exp\{{\cal G}(\as,\tL)\}$ of Eq.~(\ref{wtilde}) is singular at very large
values of $b$. The singularity occurs in the region 
$b \gtap 1/\Lambda_{\rm QCD}$, where $\Lambda_{\rm QCD}$ is the momentum scale
of the Landau pole of the perturbative running coupling $\as(\mu^2)$.
This singularity is the `perturbative' signal of the onset of non-perturbative
(NP) phenomena at very large values of $b$ 
(which practically affect the region of very small transverse momenta).
In this region NP effects cannot any longer be
regarded as small quantitative corrections and they have to be taken into
account in QCD calculations. A simple and customary procedure to include NP
effects is as follows. The singular behaviour of the perturbative form factor
$\exp\{{\cal G}(\as,\tL)\}$ is removed by using a regularization
procedure\footnote{We recall that the resummed form factor
$\exp\{{\cal G}(\as,\tL)\}$ produces a strong suppression
(${\cal G}(\as,\tL) \propto -\as \tL^2$) in the large-$b$ region where
$\as \tL^2 \gtap {\cal O}(1)$. Therefore, the choice of different regularization
procedures mildly affects \cite{Dokshitzer:hw,Collins:va,Collins:1984kg}
the results since its effects are relevant
only in the region  $b \sim {\cal O}(1/\Lambda_{\rm QCD})$
where the $b$ integral
is strongly damped by the form factor.}
and the resummed expression in Eq.~(\ref{wtilde}) is then multiplied
by a NP form factor and it is inserted as integrand of the $b$ space integral
in Eq.~(\ref{resw}).
The regularization procedure that was used in the 
{\tt DYqT} calculation \cite{Bozzi:2010xn} is the `minimal prescription' of 
Ref.~\cite{Laenen:2000de,Kulesza:2002rh}, which basically amounts to avoid the
singularity of $\exp\{{\cal G}(\as,\tL)\}$ by deforming the integration contour
of Eq.~(\ref{resw}) in the complex $b$ plane. In the {\tt DYRes} calculation of
the present work, we use a different regularization procedure by freezing the 
$b$ dependence of $\exp\{{\cal G}(\as,\tL)\}$ before reaching its singular
point. The freezing procedure follows the `$b_*$ prescription'
of Refs.~\cite{Collins:va,Collins:1984kg} and it is obtained by performing
the replacement
\begin{equation}
\label{bstar}
b^2 \to b_*^2 = b^2 \;b_{\rm lim}^2/( b^2 + b_{\rm lim}^2) 
\end{equation}
in the $b$ dependence of ${\cal G}(\as,\tL)$. The value of the parameter 
$b_{\rm lim}$ has to be large ($b_{\rm lim} M \sim b_{\rm lim} Q~\gg~1$) but smaller than the
value of $b$ at which the singularity of $\exp\{{\cal G}(\as,\tL)\}$ takes
place 
(note that the replacement in Eq.~(\ref{bstar}) has a negligible effect at
small and intermediate values of $b$ since 
$b_*^2 = b^2(1+{\cal O}(b^2Q^2/b_{\rm lim}^2Q^2)) \simeq b^2$
if $bQ~\ltap~1$).
The use of the $b_*$ freezing procedure improves the
(numerical) performances of the {\tt DYRes} code.
Additional comments on NP effects are presented in Sect.~\ref{sec:inclu}.

\section{Numerical results at the LHC}
\label{sec:results}

In this Section 
we consider the processes $pp\to Z/\gamma^*\to l^+l^-$ and $pp\to W^\pm\to l\nu_l$ at LHC energies.
We present our resummed results at NNLL+NNLO and NLL+NLO accuracy,
and we compare them  with some of the available LHC data.
We compute the hadronic cross sections at NNLL+NNLO (NLL+NLO) accuracy by using  
the NNPDF3.0 NNLO (NLO) \cite{Ball:2014uwa} parton densities
functions (PDFs),
with $\as(m_Z^2)=0.118$ and with
$\as(\mu_R^2)$ evaluated at 3-loop (2-loop) order.
As in customary fixed-order calculations at high invariant mass 
($M={\cal O}(m_Z))$, we consider $N_f=5$ flavours of light quarks and we treat
them in the massless approximation.

As for the 
EW couplings, we use the so called $G_\mu$ scheme,
where the input parameters are $G_F$, $m_Z$, $m_W$.
In particular, we use the PDG 2014 \cite{Agashe:2014kda} values
$G_F = 1.1663787\times 10^{-5}$~GeV$^{-2}$,
$m_Z = 91.1876$~GeV, $\Gamma_Z=2.4952$~GeV, $m_W = 80.385$~GeV and $\Gamma_W=2.085$~GeV and
in the case of $W^\pm$ production, we use the (unitarity constrained) 
CKM matrix elements 
$V_{ud} = 0.97427$, $V_{us} = 0.22536$,
$V_{ub} = 0.00355$, $V_{cd} = 0.22522$, $V_{cs} = 0.97343$, $V_{cb} = 0.0414$.
Our calculation implements 
the leptonic decays $Z/\gamma^* \to l^+l^-$ and $W\to l\nu_l$ 
(we include the effects of the $Z/\gamma^*$ interference and of the finite width
of the $W$ and $Z$ bosons)
with
the corresponding spin correlations and the full dependence  
on the kinematical variables of final state leptons.
This allows us to take into account the typical 
kinematical cuts on final state leptons that are
considered in the experimental analysis. 
As discussed in Sect.~\ref{sec:theory}, the resummed calculation 
at fixed lepton momenta requires a $q_T$-recoil procedure.
We implement a procedure that is described in Appendix~\ref{app}, and that is practically
equivalent to compute the Born level distribution 
$d{\hat \sigma}^{(0)}/d{\bomega}$ of Eqs.~(\ref{resum}) and (\ref{Frecoil})
in the Collins--Soper rest frame \cite{Collins:1977iv} (this is exactly the same procedure as
used in other resummed calculations 
\cite{Balazs:1995nz,Balazs:1997xd,Ellis:1997sc,Cieri:2015rqa}).
As explained in Sect.~\ref{sec:theory}, 
the {\tt DYRes} resummed calculation uses a smooth switching procedure 
(see Eq.~(\ref{switch})) in the large-$q_T$ region.
In our numerical implementation the parameters in Eq.~(\ref{fswitch}) are
chosen to be $\Delta q_T=M/(2\sqrt{2})$ and $q_T^{\rm sw.}=3M/4$.

\subsection{Inclusive results at fixed $q_T$}
\label{sec:inclu}

We start the presentation of our results by discussing
some general features of the $q_T$ spectrum that can be addressed
at the inclusive level, i.e. after integration over the lepton angular variables
$\bomega$ and over the rapidity $y$ of the lepton pair.
Unless otherwise specified, the numerical results of this Subsection are
obtained by using the code {\tt DYqT} \cite{Bozzi:2008bb,Bozzi:2010xn}.
The code {\tt DYqT} is publicly available and it can be downloaded from the URL
address {\tt http://pcteserver.mi.infn.it/\char`\~ferrera/dyqt.html}~.

We first consider the dependence on the auxiliary scales
$\mu_F$, $\mu_R$ and $Q$. These scales have to be set at values of the order of
the invariant mass $M$ of the produced system, with no definite preference
for specific values. Then, scale variations around the chosen central value
can be used to estimate the size of yet uncalculated higher-order terms
and the ensuing perturbative uncertainties.
In the NNLL+NNLO studies of
Refs.~\cite{Bozzi:2005wk,deFlorian:2011xf,deFlorian:2012mx} on Higgs boson production, and in our previous work on vector boson production \cite{Bozzi:2010xn} the central
reference values of the scales were chosen as $\mu_F=\mu_R=2Q=m_F$, where $m_F$
is the mass of the produced boson (the Higgs boson mass in the case of Higgs boson
production, and the vector boson mass in the case of vector boson production). 
In the case of Higgs boson production, this choice gives consistent 
NLL+NLO and NNLL+NNLO results with a reduced scale dependence at NNLL+NNLO 
level and, in particular,
with a nice overlap of the NLL+NLO and NNLL+NNLO uncertainty bands (see, e.g., Fig.~2 of Ref.~\cite{deFlorian:2011xf}).
In the case of vector boson production, our previous studies were focused on
Tevatron energies, and a similar pattern was observed \cite{Bozzi:2010xn}. When moving to 
vector boson production at
LHC energies, we notice that the factorization-scale
dependence exhibits a (slightly) different behaviour.

\begin{figure}[th]
\centering
\hspace*{-0.8cm}
\subfigure[]{
\includegraphics[width=3.53in]{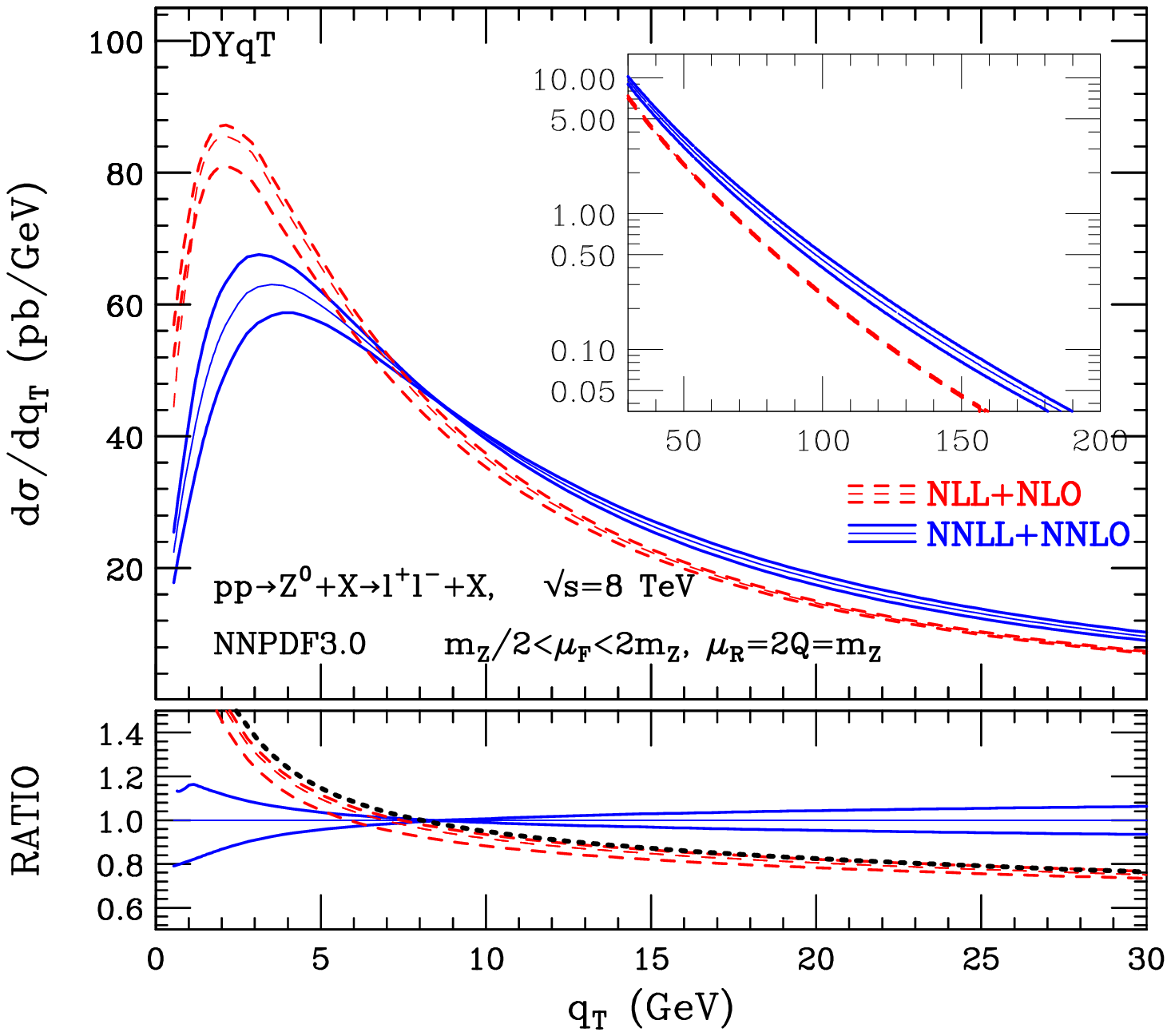}
}
\hspace*{.15cm}
\subfigure[]{
\includegraphics[width=3.28in]{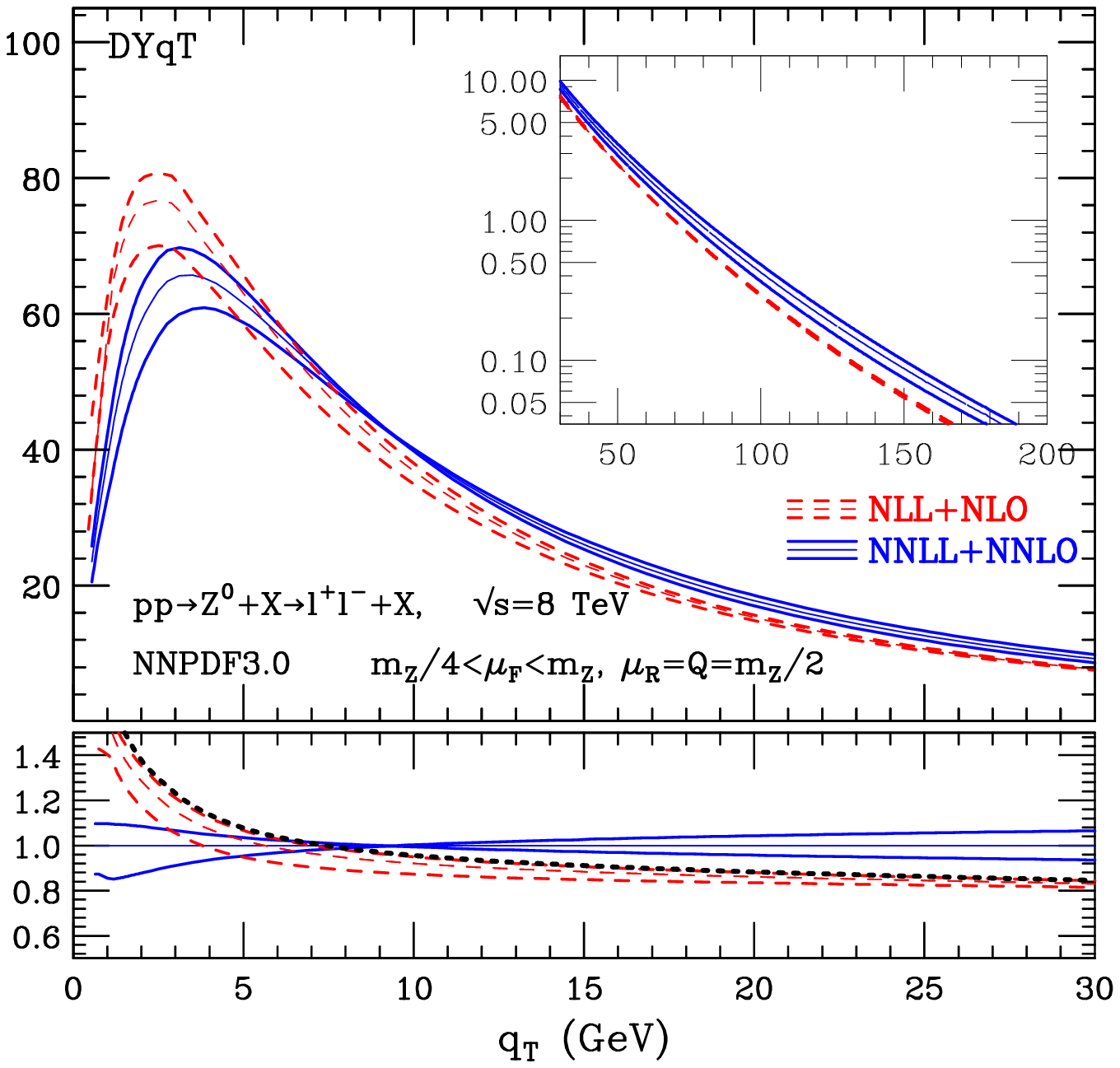}}
\caption{\label{fig:dyqtmuF}
{\em The $q_T$ spectrum of $Z$ bosons at  the LHC ($\sqrt{s}=8$~TeV). 
The bands of the NLL+NLO (red dashed) and NNLL+NNLO (blue solid) results
are obtained by performing $\mu_F$ variations 
around the central values
$\mu_F=m_Z$ (left panel) and $\mu_F=m_Z/2$ (right panel).
The lower panel presents the ratio of the NLL+NLO and NNLL+NNLO results
with respect to the NNLL+NNLO result at the central value of $\mu_F$.
The result of the convolution of the NLL+NLO partonic cross section with
NNLO PDFs at the central value of $\mu_F$ is also reported (black dotted)
in the lower panel.
}}
\end{figure}

In Fig.~\ref{fig:dyqtmuF} we present results for the $q_T$ spectrum of on-shell $Z$ 
bosons produced at the LHC ($\sqrt{s}=8$~TeV) and the corresponding dependence
on factorization-scale variations.
In the left panel the central scale is $\mu_F=\mu_R=2Q=m_Z$, while in the 
right panel
the central scale is $\mu_F=\mu_R=Q=m_Z/2$.
In both panels, we present the NLL+NLO and NNLL+NNLO results at the central
scale and corresponding bands that are obtained by varying (up and down)
the factorization scale by a factor of 2 around its corresponding central value.
The lower panel in Figs.~\ref{fig:dyqtmuF}(a) and \ref{fig:dyqtmuF}(b)
presents the ratio of the various results with respect to the NNLL+NNLO result
at the corresponding central scale.
If $\mu_F=\mu_R=2Q=m_Z$ is the central scale choice (left panel),
we see that the factorization-scale bands at NLL+NLO and NNLL+NNLO accuracy
never overlap, except for the tiny region
around $q_T\sim 7$~GeV where they cross each other. The lack of overlap
is particularly evident
in the peak region, where the bulk of the events is placed: here the central 
NLL+NLO and NNLL+NNLO results differ by about $30$\%.
We also notice that throughout the region of small and intermediate values of  
$q_T$ ($q_T \ltap$30~GeV)
the size of the NLL+NLO band is rather small
and it is always (with the exception of a small region around
$q_T\sim 8$~GeV) smaller than the size of the NNLL+NNLO band, and this suggests
that an accidental cancellation of the $\mu_F$ dependence may occur at the
NLL+NLO level with this choice of central scale.
In the right panel we observe a $\mu_F$-dependence behaviour that is 
qualitatively similar but quantitatively different from that in the left panel.
If $\mu_F=\mu_R=Q=m_Z/2$ is the central scale choice (right panel),
the NLL+NLO and NNLL+NNLO bands are closer and they overlap at small 
transverse momenta. The overlap occurs in a limited region of $q_T$ that,
nevertheless, includes the peak region.
The shape of the spectra
appears closer when going from NLL+NLO to NNLL+NNLO accuracy,
and the NLL+NLO band is wider than the NNLL+NNLO one in
the small and intermediate region of $q_T$.
Note that the central values of $\mu_R$ are $\mu_R=m_Z$ and $\mu_R=m_Z/2$
in the left and right plot, respectively, but we have checked that this
difference has little effect: the observed different behaviour is mainly due
to the different central value of $\mu_F$.
In summary, the $\mu_F$ dependence observed in the left panel of 
Fig.~\ref{fig:dyqtmuF} suggests that the corresponding scale variation bands
(and especially the NLL+NLO band)
are likely to underestimate the perturbative uncertainties of the calculation. 
Based on these observations, in the rest of the paper, we will adopt
$\mu_F=\mu_R=Q=m_Z/2$ as reference values of the central scales.

The differences between the NLL+NLO and NNLL+NNLO results have a twofold origin.
Part of the differences is due to the next-order radiative corrections in the
partonic cross sections, and the remaining part is due to the increased order of
the PDFs. To quantify the impact of these two different contributions, we have
considered the result that is obtained by convoluting the NLL+NLO partonic cross
sections with the NNLO PDFs. This result, at the central scales that are
considered in Figs.~\ref{fig:dyqtmuF}(a) and
\ref{fig:dyqtmuF}(b), is reported (see
the black dotted line) in the corresponding lower panel, 
and we can see that it is quite close to the NLL+NLO
result with NLO PDFs. In other words, a large part of the quantitative
differences between the NLL+NLO and NNLL+NNLO results is due to the 
corresponding differences at the level of the partonic cross sections.

\begin{figure}[th]
\centering
\hspace*{-0.8cm}
\subfigure[]{
\includegraphics[width=3.53in]{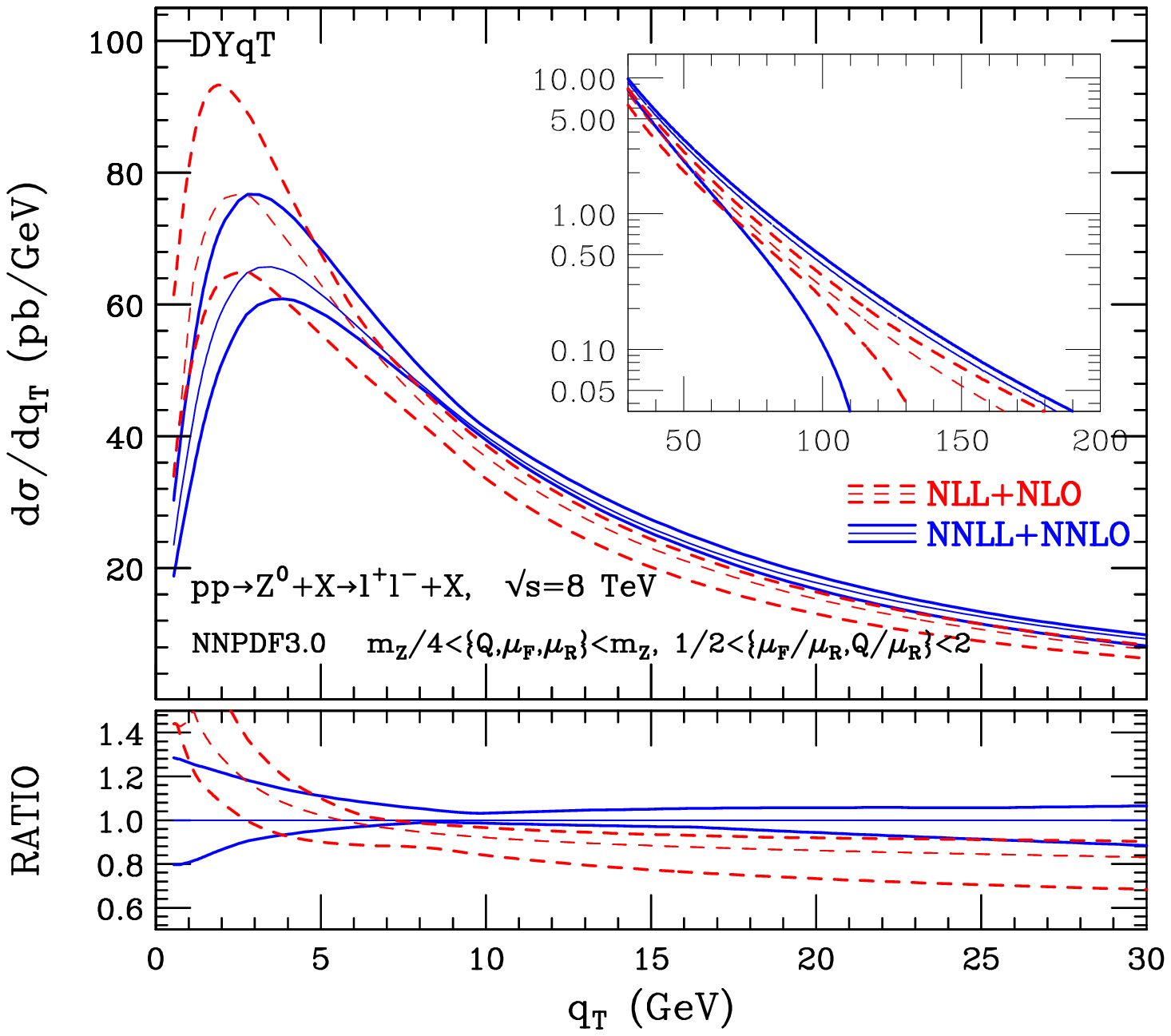}
}
\hspace*{.15cm}
\subfigure[]{
\includegraphics[width=3.28in]{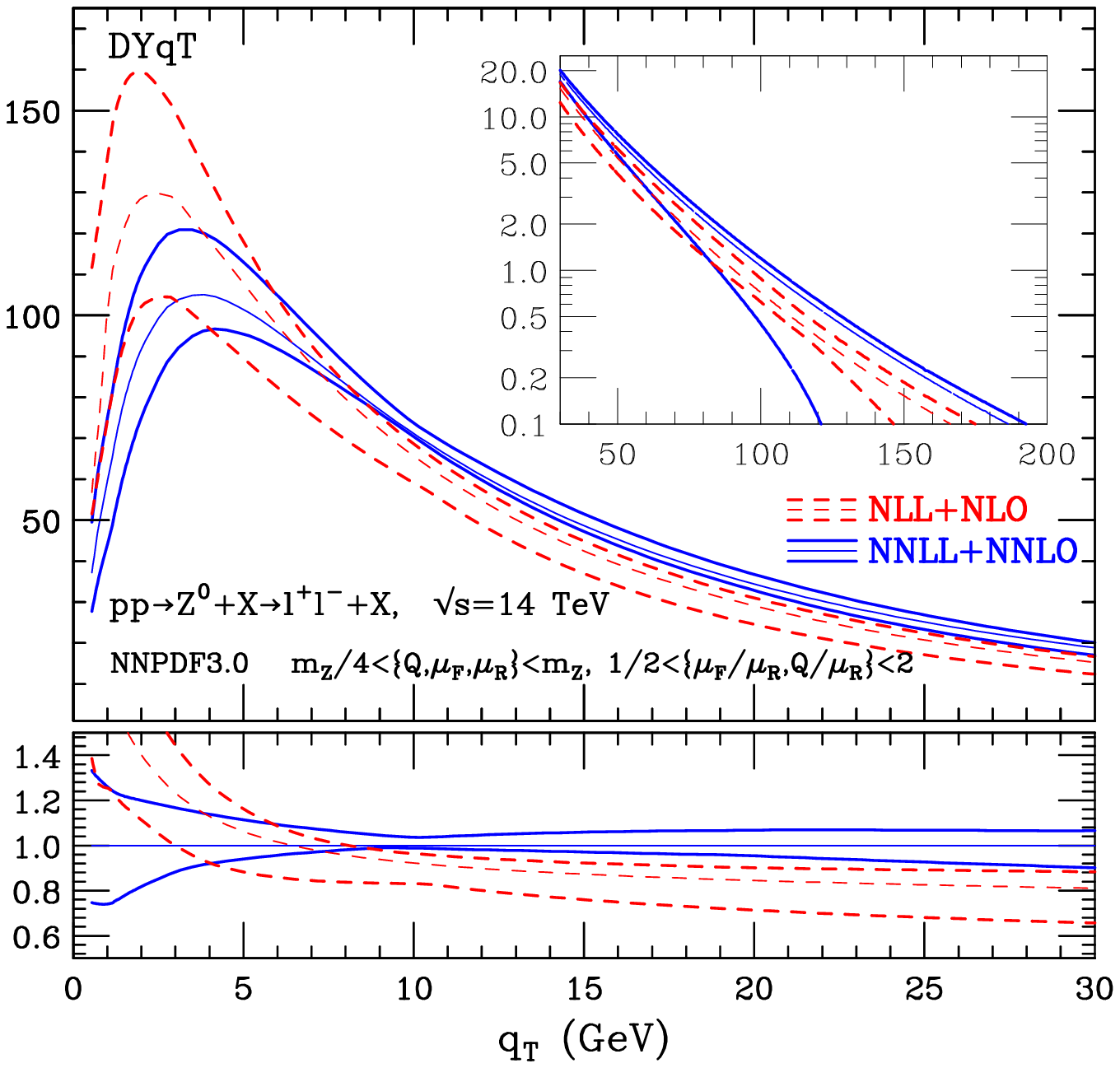}}
\caption{\label{fig:dyqt}
{\em  NLL+NLO (red dashed) and NNLL+NNLO (blue solid) results for
the $q_T$ spectrum of $Z$ bosons at the LHC with energies
$\sqrt{s}=$8~TeV (left panel)
and $\sqrt{s}=$14~TeV  (right panel). 
The NLL+NLO and NNLL+NNLO results with the central value
$\mu_F=\mu_R=Q=m_Z/2$ of the scales are enclosed by
corresponding bands.
The bands are obtained by performing $\{\mu_F, \mu_R, Q \}$ variations
(as described in the text) around the central value $m_Z/2$.
The lower panel presents the ratio of the scale-dependent NLL+NLO and NNLL+NNLO  
results with respect to the NNLL+NNLO result at the central value of the scales.
}}
\end{figure}

The NLL+NLO and NNLL+NNLO results for the $q_T$ spectrum of on-shell
$Z$ boson produced at the LHC with different collision energies
are presented in Fig.~\ref{fig:dyqt}.
We consider two centre--of--mass energies: $\sqrt{s}=$8~TeV (Fig.~\ref{fig:dyqt}
left) and $\sqrt{s}=$14~TeV (Fig.~\ref{fig:dyqt} right).
At each logarithmic accuracy we present the result at the central value
$\mu_F=\mu_R=Q=m_Z/2$ of the scales and a corresponding band.
The bands provide an estimate of the perturbative uncertainties of the
calculations due to missing higher-order contributions.
The bands are obtained
through independent variations of $\mu_F$, $\mu_R$ and $Q$
by following the procedure of Ref.~\cite{Bozzi:2010xn}:
we independently vary
$\mu_F$, $\mu_R$ and $Q$ in the range $m_Z/4\leq \{\mu_F, \mu_R, Q\}\leq m_Z$ with the constraints
$0.5\leq \mu_F/\mu_R\leq 2$ and $0.5\leq Q/\mu_R\leq 2$.
We remind the reader that the constraint on $\mu_F/\mu_R$ is introduced to avoid large
logarithmic contributions ($\ln(\mu_F/\mu_R)$ terms
from the evolution of the parton densities) in the perturbative expansion
of the hard/collinear factor ${\cal H}_V$ of Eq.~(\ref{wtilde}).
Analogously, the constraint on $Q/\mu_R$
avoids large logarithmic terms ($\ln(Q/\mu_R)$)
in the resummed expansion of the form factor $\exp\{{\cal G}\}$
of Eq.~(\ref{wtilde}).
The lower panels in Fig.~\ref{fig:dyqt} present the ratio of the scale-dependent
NLL+NLO and NNLL+NNLO results 
with respect to the NNLL+NNLO result at the central value $\mu_F=\mu_R=Q=m_Z/2$
of the scales.

The region of small and intermediate values of $q_T$ is shown in the main 
panels of Fig.~\ref{fig:dyqt}.
At fixed centre--of--mass energy the NNLL+NNLO $q_T$ spectrum is harder than
the spectrum at NLL+NLO accuracy. At fixed value of $q_T$ the cross section
sizeably increases by increasing the centre--of--mass energy from 8~TeV to
14~TeV. The shape of the NNLL+NNLO $q_T$ spectrum is slightly harder at the
higher energy.
The NLL+NLO scale-variation band is wider than the NNLL+NNLO band.
The NLL+NLO and NNLL+NNLO bands overlap at small transverse momenta and remain
very close by increasing $q_T$ (the differences with respect to the plot
on the right-hand side of Fig.~\ref{fig:dyqtmuF} are due to the additional
dependence on $\mu_R$ and $Q$).
The NNLL+NNLO (NLL+NLO) scale dependence is about $\pm 10$\% ($\pm 20$\%) 
at the peak, it decreases to about
$\pm 2$\% ($\pm 7$\%) at $q_T \simeq 10$~GeV and increases to about 
$\pm 6$\% ($\pm 10$\%) at $q_T\sim 25$~GeV.
Since the NNLL+NNLO and NLL+NLO bands do not exactly touch each other in the
region where $q_T \gtap 8$~GeV, one may argue that the `true' perturbative
uncertainty  of the NNLL+NNLO result in this region is slightly larger than the
size of the NNLL+NNLO scale dependence band (for instance, one may use
\cite{Bozzi:2008bb} the difference between the NNLL+NNLO central scale result
and the upper
line of the NLL+NLO band in Fig.~\ref{fig:dyqt} to estimate the uncertainty
of the NNLL+NNLO result).

The inset plots show the cross section in the large-$q_T$ region.
The resummation results obtained with {\tt DYqT} and reported in the inset
plots are presented for completeness
and mainly for illustrative purposes. 
At large values of $q_T$ ($q_T\gtap m_Z$) the resummed result looses predictivity, and its perturbative
uncertainty becomes large. In this region of transverse momenta
we see that the
uncertainty band increases in going from the NLL+NLO to the NNLL+NNLO level.
However, as already mentioned in Sect.~\ref{sec:theory}, 
at high $q_T$ the resummation cannot improve
the predictivity of fixed-order calculations and 
the {\tt DYqT} result in Fig.~\ref{fig:dyqt} cannot be regarded as reference
theoretical result. 
The resummed result has to be replaced by the standard fixed-order prediction.
The NNLO (NLO) result (which is not shown in Fig.~\ref{fig:dyqt})
lies inside the NNLL+NNLO (NLL+NLO)
band and the former has a smaller scale dependence than the latter.
We also note that, at high  $q_T$, the preferred reference central scales 
$\mu_R$ and
$\mu_F$ of the fixed-order prediction should be of the order of
$\sqrt{ m_Z^2 + q_T^2}$ (rather than of the order of $m_Z$).

We also recall that, increasing $q_T$ throughout the high-$q_T$ region,
fixed-order QCD calculations are affected by additional and potentially-large
logarithmic terms. These are collinear (fragmentation) contributions
\cite{Berger:2001wr,Berger:2015nsa}, which become more relevant by increasing
the ratio $q_T/M$, and soft (threshold) contributions
\cite{Kidonakis:1999ur,Becher:2011fc}, which become more relevant by increasing
the ratio $q_T/\sqrt s$ (or $q_T/\sqrt{\hat s}$).

We have so far discussed only uncertainties from missing 
higher-order contributions. Before moving to consider the case in which cuts on
the final-state leptons are applied, we briefly discuss two additional sources
of QCD uncertainties on the $q_T$ spectrum:
the uncertainty from PDFs and that from NP effects. 
We consider these effects in turn.

Modern sets of PDFs include an estimate of the errors (mainly experimental
errors) in their determination from global data fits, and this estimate can
then be used to compute the ensuing PDF uncertainty on the QCD calculation
of hadron collider observables.
In Fig.~\ref{fig:pdfs} we consider $Z$ boson production at NNLL+NNLO accuracy.
In Fig.~\ref{fig:pdfs}~(a) we report the NNLL+NNLO results of 
Fig.~\ref{fig:dyqt}~(a) ($\sqrt{s}=$14~TeV) and the effect of the 
PDF uncertainty at 68\% CL on the NNLL+NNLO calculation at the central scale
value $\mu_F=\mu_R=Q=m_Z/2$. In Fig.~\ref{fig:pdfs}~(b) the
scale-dependence and PDF-uncertainty bands are normalized 
to the central NNLL+NNLO prediction, and we present results at both energies
$\sqrt{s}=$8~TeV (lower panel) and $\sqrt{s}=$14~TeV (upper panel).
We see that the PDF uncertainty is smaller than the scale uncertainty.
Moreover, the PDF uncertainty is approximately independent 
on the transverse momentum, 
and it has a value of about $\pm 3$\% at
both energies $\sqrt{s}=8$~TeV and $\sqrt{s}=14$~TeV.

NP effects are known 
to increasingly affect the transverse-momentum spectrum as $q_T$ decreases
towards $q_T\to 0$.
A detailed study of these effects is beyond the scope of the present work.
We limit ourselves to roughly estimate the possible impact of such effects,
and we use a very simple model in which the perturbative form factor 
$\exp\{{\cal G}(\as,\tL)$ in Eq.~(\ref{wtilde})
is multiplied by a NP form factor $S_{NP}(b)=\exp\{-g_{NP}b^2\}$, which
produces a Gaussian smearing of the $q_T$ distribution at small-$q_T$ values.
We vary the value of the parameter $g_{NP}$ in a quite wide (`conservative')
range, $0\leq g_{NP}\leq 1.2$~GeV$^2$,  and in Fig.~\ref{fig:pdfs}(a)
(black band) we show the ensuing quantitative effects on the $q_T$ spectrum.
In Fig.~\ref{fig:pdfs}(b) the NP effects are normalized with respect to the
perturbative NNLL+NNLO result at central value of the scales.

Comparing the lower panels of Fig.~\ref{fig:dyqt} with Fig.~\ref{fig:pdfs}, we
can first make an overall qualitative comment. Perturbative corrections make the
$q_T$ spectrum harder in going from NLL+NLO to NNLL+NNLO accuracy, and this
occurs at both small and intermediate values of $q_T$. NP effects increase the
hardness of the $q_T$ spectrum at small values of $q_T$ and they are negligible 
at intermediate values of $q_T$. Therefore, we note a non trivial interplay
of perturbative and NP effects. In particular, at small values of $q_T$
higher-order perturbative contributions can be mimicked by NP effects.

At the quantitative level, in Fig.~\ref{fig:pdfs} we see
that the NNLL+NNLO result supplemented with NP effects is very close
to the perturbative result except in the very low $q_T$ region 
($q_T\ltap 3$ GeV), i.e. below the peak of the $q_T$ distribution.
In the region $3~{\rm GeV} \ltap q_T \ltap 10$~GeV, the size of the NP band is
similar to that of the PDF uncertainty band. At larger values of $q_T$,
the NP effects vanish (the size of the NP band is smaller than about 
2\% starting from $q_T \sim 15$~GeV).

We note that our simple model treats the regularization of the perturbative form
factor (through the `minimal prescription', see Sect.~\ref{sec:theory}) and the
NP form factor in an uncorrelated way, and this produces a conservative estimate
of NP uncertainties. In other words, the model underestimates the potential
of the resummed calculation at very small values of $q_T$. For instance, the NP
model can be improved by correlating the interplay between the perturbative form
factor (and, e.g., its scale variation dependence) and the NP form factor
(and the value of $g_{NP}$), and further constraints on the NP model can be
possibly obtained by inputs from comparisons with experimental data.

In summary, from our brief discussion on the possible impact of NP effects
for vector boson production at the LHC, we conclude that our conservative
estimate leads to quantitative effects that are small and well within the scale
variation dependence, still in the very low $q_T$ region. A quantitatively 
similar conclusion applies to the effect of PDF uncertainties.
Based on these observations (and for practical purposes), 
in the presentation of our results of Sect.~\ref{sec:lepton}
we limit ourselves to considering only the perturbative calculation and the
corresponding scale variation uncertainties.

\begin{figure}[th]
\centering
\hspace*{-0.8cm}
\subfigure[]{
\includegraphics[width=3.4in]{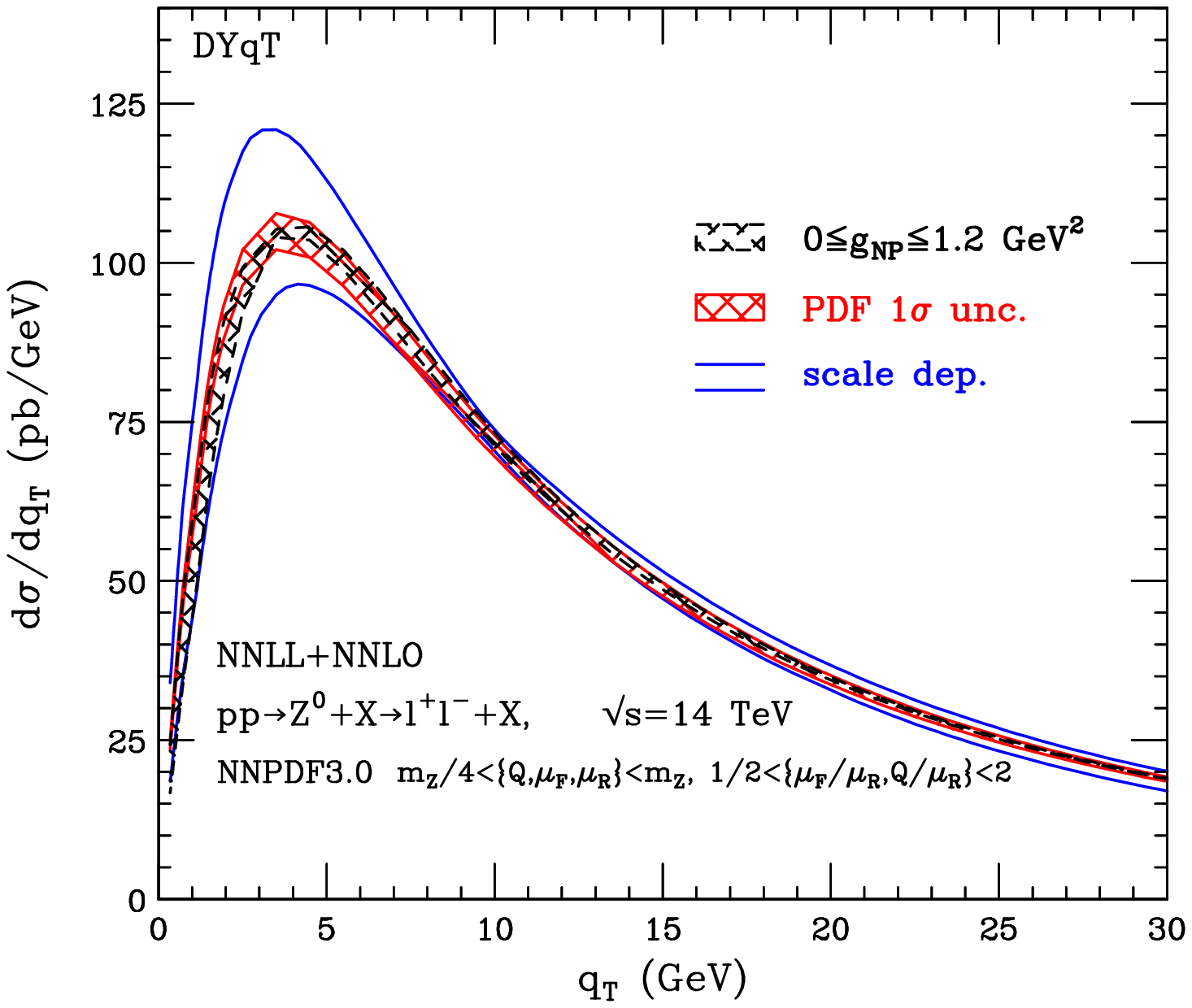}
}
\hspace*{.5mm}
\subfigure[]{
\includegraphics[width=3.4in]{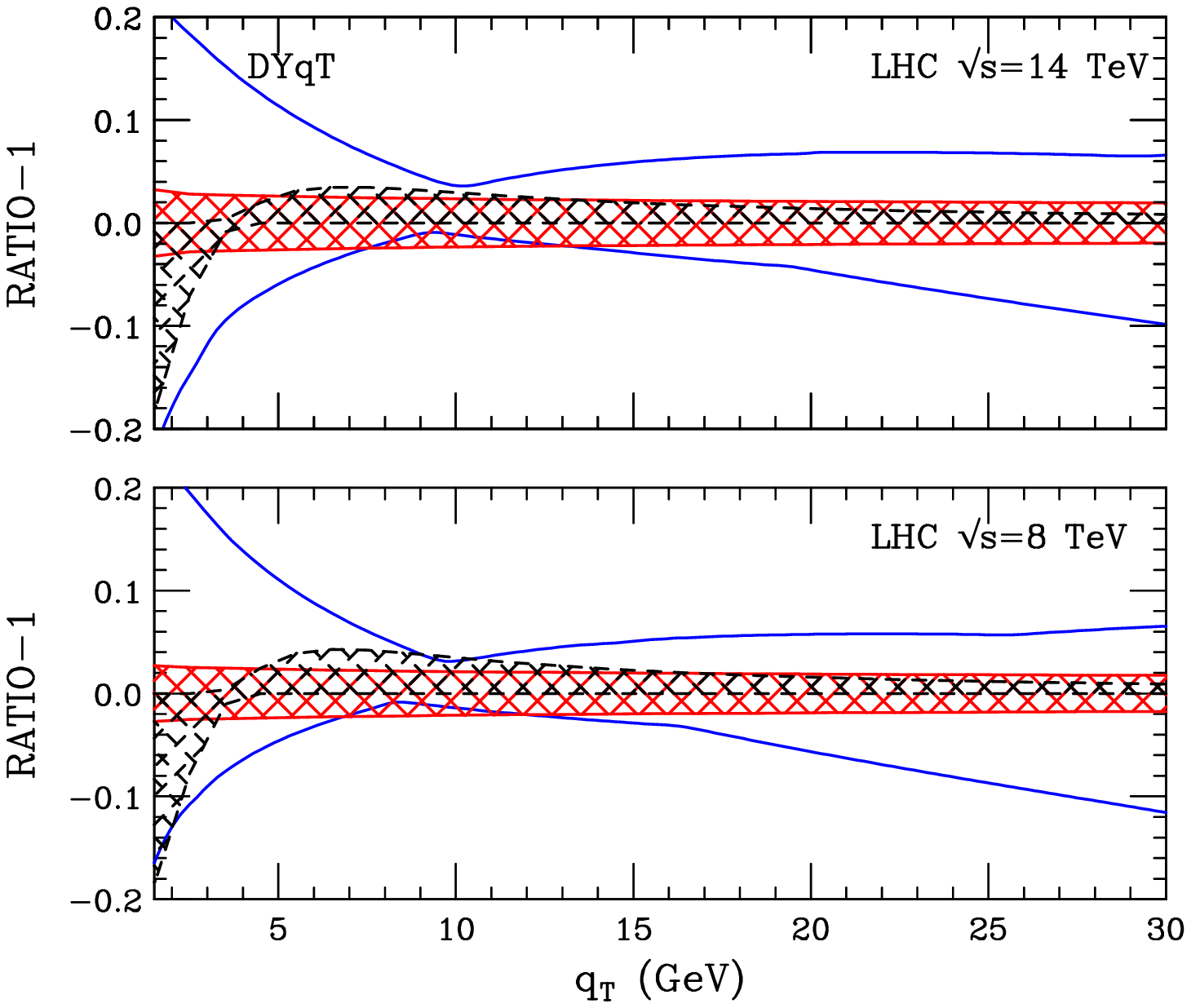}}
\caption{\label{fig:pdfs}
{\em (a) The $q_T$ spectrum at NNLL+NNLO accuracy
for $Z$ boson production at the LHC  with $\sqrt{s}=$14~TeV. 
Comparison of scale dependence (blue solid)
and PDF (red crossed solid) uncertainties. The possible impact of NP effects 
is also shown (black crossed dashed).
(b) The same results are normalized to the central NNLL+NNLO prediction at
$\sqrt{s}=14$~TeV (upper panel), and corresponding results are shown at  
$\sqrt{s}=8$~TeV (lower panel).
}}
\end{figure}

We conclude this subsection by presenting a comparison between the 
{\tt DYqT} results and the results of the `multidifferential' program 
{\tt DYRes}.
When no cuts are applied on the final-state leptons, 
the $q_T$ spectrum of the on-shell vector 
boson obtained with {\tt DYRes} has to be in agreement with the one obtained with 
the numerical program {\tt DYqT}. We have numerically checked that  
this is indeed the case. For illustrative purposes, we show the results
of a comparison  in Fig.~\ref{fig:pTZ}. Here we consider the $q_T$ spectrum for
on-shell $Z$ boson production at the LHC with $\sqrt{s}=7$~TeV.
The {\tt DYqT} (solid line) and {\tt DYRes} (histogram) results at central value
of the scales are compared at both NLL+NLO (red) and NNLL+NNLO (blue) accuracy.
At small and intermediate values of $q_T$ (main plot in Fig.~\ref{fig:pTZ}),
the {\tt DYqT} and {\tt DYRes} results agree 
(within the statistical uncertainties of the {\tt DYRes} 
code\footnote{Here and in the following the errors reported in the tables 
and on the histograms refer to a numerical estimate of the accuracy of 
the Monte Carlo integration in the {\tt DYNNLO} and {\tt DYRes} codes.})
at both level of logarithmic accuracy.
The quantitative degree of agreement is more clearly visible in the lower panel,
which presents the result of the calculation of the binned
ratio between the {\tt DYRes} and {\tt DYqT} results at both NLL+NLO and 
NNLL+NNLO accuracy. The ratio is everywhere consistent with unity within the
numerical accuracy of its computation (the numerical errors in the computation
of the binned ratio are below about 1\% at small values of $q_T$, and they are
still below about 2\% in the region $30~{\rm GeV} \ltap q_T \ltap 50~{\rm GeV}$
where the value of the cross section sizeably decreases).

\begin{figure}[!ht]
\centering
\includegraphics[width=0.9\textwidth]{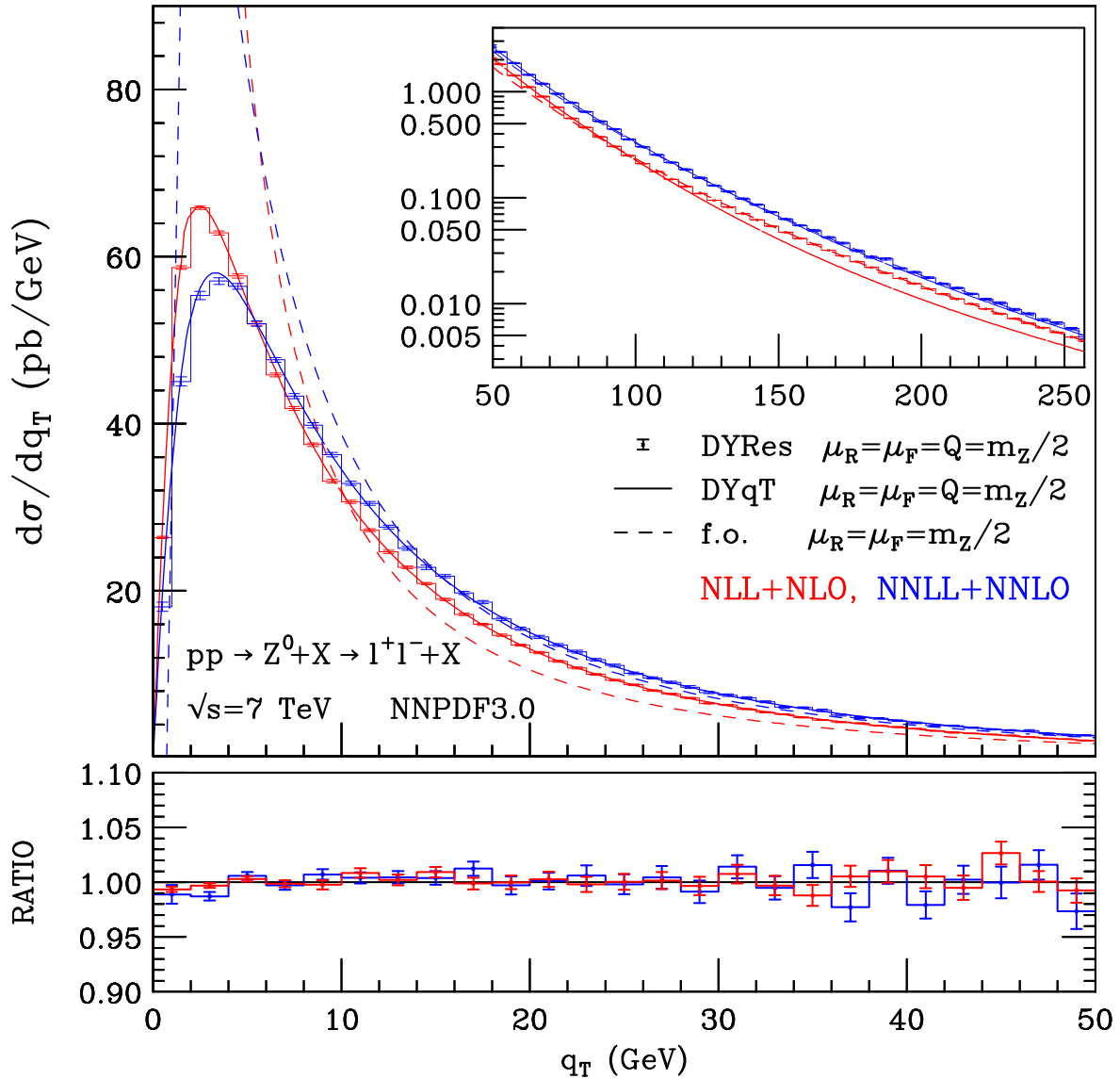}
\caption{{\em The $q_T$ spectrum of on-shell $Z$ bosons at the 
LHC ($\sqrt{s}=7$~TeV). Comparison of the {\tt DYRes} (histograms) and 
{\tt DYqT} (solid lines) 
resummed results at NLL+NLO (red) and NNLL+NNLO (blue) accuracy.
The corresponding fixed-order calculations (dashed lines) at 
${\cal O}(\as)$ (red) and ${\cal O}(\as^2)$ (blue)
are also shown.
The ratio between the {\tt DYRes} and {\tt DYqT} results is shown in the lower
panel.
}}
\label{fig:pTZ}
\end{figure}

We recall (see Sect.~\ref{sec:theory}) that, at the inclusive level,
the {\tt DYqT} and {\tt DYRes} calculations involve differences in the numerical
implementation and two additional differences related to the treatment of the
very low $q_T$ and high-$q_T$ regions.
At very low values of $q_T$, the difference is due to the regularization
procedure of the perturbative form factor for very large values of the impact
parameter $b$: the {\tt DYqT} calculation uses the `minimal prescription', while
the {\tt DYRes} calculation uses the $b_*$ freezing procedure. In our actual
calculation with {\tt DYRes} the value of $b_{\rm lim}$ in Eq.~(\ref{bstar})
is set to $b_{\rm lim} = b_{\rm max}$, where $b_{\rm max}$ is the maximum value
of $b$ that can be reached before encountering the singularity of the 
perturbative form factor (setting $b_{\rm lim} = b_{\rm max}$ we do not
introduce any additional regularization parameter, analogously to the case of
the `minimal prescription').
The 
value of $b_{\rm max}$ depends on the renormalization and resummation 
scales $\mu_R$ and $Q$ and,  
in the case of $Z$ and $W$ production around the central value of the scales, 
the typical value is 
$b_{\rm max} Q \sim 1.2 \cdot 10^3 \mu_R/m_Z$.
We have checked that the `minimal prescription' and the choice
$b_{\rm lim} = b_{\rm max}$ give basically the same numerical results, also at
very small values of  $q_T$ ($q_T \sim 1$~GeV).
This numerical agreement is also visible (lower panel in Fig.~\ref{fig:pTZ})
from the ratio between the {\tt DYRes} and {\tt DYqT} results at low values of 
$q_T$.

At large values of $q_T$, the {\tt DYRes} calculation implements the smooth
switching procedure of Eqs.~(\ref{switch})--(\ref{fswitch}). 
The large-$q_T$ region is shown in the inset plot of Fig.~\ref{fig:pTZ}, and
here the differences between the {\tt DYqT} and {\tt DYRes} calculations are due
to the smooth switching procedure. In the high-$q_T$ region the {\tt DYRes}
result at NNLL+NNLO (NLL+NLO) accuracy 
basically agrees with the customary fixed-order result at 
${\cal O}(\as^2)$ (${\cal O}(\as)$). The differences between the {\tt DYRes}
and {\tt DYqT} results (consistently) decrease in going from NLL+NLO
to NNLL+NNLO accuracy, and they are small at the NNLL+NNLO level.
At both level of logarithmic accuracy, the {\tt DYRes}
and {\tt DYqT} results agree within their corresponding scale variation
uncertainties (which are not shown in the inset plot), and the {\tt DYRes}
result has a reduced scale dependence (it matches the scale dependence of the
corresponding fixed-order result). The introduction of the 
smooth switching procedure in the {\tt DYRes} calculation has practically
a negligible quantitative effect on the unitarity constraint that is fulfilled
by the {\tt DYqT} calculation.
In Table~\ref{tab:inclusive} we report the total cross sections for both
$Z$ and $W$ production at $\sqrt{s}=7$~TeV, and we compare the resummed 
{\tt DYRes} results with the corresponding fixed-order results obtained with the
{\tt DYNNLO} code. We see
that the NLL+NLO (NNLL+NNLO) total cross section agrees 
with the NLO (NNLO) result to better than 1\% accuracy.

\begin{table}[!ht]
\begin{center}
\begin{tabular}{|c||c|c|c|c|}
\hline
Cross section [pb] & NLO & NLL+NLO & NNLO & NNLL+NNLO\\ \hline \hline
$pp\to Z \to l^+l^-$ & $904.3\pm 0.2$ & $904.6\pm 0.4$ &  $949.1\pm 0.7$  & $947.3\pm 0.9$ \\ \hline
$pp\to W^{(\pm)}\to l^{(\pm)}\nu$ & $9819\pm 2$  &  $9813\pm 4$  & $10337\pm 6$  & $10328 \pm 9$ \\ \hline
\end{tabular}
\end{center}
\caption{{\em Total cross sections at the LHC ($\sqrt{s}=7$~TeV): fixed-order
results and corresponding resummation results of the {\tt DYRes} numerical program.
}}
\label{tab:inclusive}
\end{table}

In the main plot of Fig.~\ref{fig:pTZ}, we also present a complementary
information on the results of the fixed-order calculations (dashed lines) at 
${\cal O}(\as)$ (red dashed) and ${\cal O}(\as^2)$ (blue dashed).
At intermediate values of $q_T$ the differences between the resummed results 
at two subsequent orders
are smaller than the differences between the corresponding
fixed-order results at two subsequent orders.
The differences between the resummed results and the corresponding fixed-order
results sizeably increase by decreasing $q_T$. 
At small values of $q_T$, the result at ${\cal O}(\as)$ increases towards 
large positive values (they are outside the vertical size of the plot) and,
in a first very small bin (not shown in the plot) around $q_T=0$, the 
${\cal O}(\as)$ result would be very large and negative.
The result at ${\cal O}(\as^2)$ has a very high unphysical peak (it is outside
the vertical size of the plot) around $q_T \sim 4$~GeV, then it decreases
towards very large negative values and,  
in a first very small bin (not shown in the plot) around $q_T=0$, 
the ${\cal O}(\as^2)$ result would be very large and positive.


\subsection{Vector boson production at the LHC}
\label{sec:lepton}


In this Section we consider ($q_T$ related) physical observables 
that depend on the individual lepton momenta and on the kinematics of the lepton
pair. The dependence can be indirect, through the application of 
acceptance cuts, and direct, through the definition of the observable.
Therefore, the resummed calculation presented in this Section are performed
by using the numerical program {\tt DYRes}.

We start our presentation by considering
the measurements of the $q_T$ spectrum of dilepton pairs at the LHC
with $\sqrt{s}=7$~TeV, 
as reported by the CMS~\cite{Chatrchyan:2011wt} and 
ATLAS~\cite{Aad:2014xaa} Collaborations
with an integrated luminosity of 36~pb$^{-1}$ and 4.7~fb$^{-1}$, respectively.
The cuts that define the {\em fiducial} region in which the measurements
are performed (our corresponding resummed calculation of the 
$Z/\gamma^*$ spectrum is carried out in the same region)
are as follows.
In the case of the CMS analysis the invariant mass $m_{ll}$ of the lepton pair
is required to be in the range 60~GeV $< m_{ll} <$ 120~GeV, and
the leptons must be in the central rapidity region, with pseudorapidity
$|\eta^l|<2.1$, and they have 
a transverse momentum $p_T^l>20$~GeV. 
In the case of the ATLAS analysis the fiducial region is defined by:  
66~GeV $< m_{ll} <$ 116~GeV, $|\eta^l|<2.4$ and $p_T^l>20$~GeV. 

The results of our resummed calculation are shown in Fig.~\ref{fig:Zqt}~(a) and (b). 
The blue-solid (red-dashed) histogram is the NNLL+NNLO (NLL+NLO) prediction for the $q_T$ spectrum, 
which is normalized to the cross section in 
the fiducial region, and the points are the data with the corresponding
experimental errors.
The inset plot shows the high-$q_T$ region.
To facilitate the 
comparison between the data and the perturbative calculation 
we consider their ratio with respect 
to a reference theoretical result. We choose the NNLL+NNLO result at 
central values of the scales ($\mu_F=\mu_R=Q=m_Z/2$) as reference result.
The lower panel shows the data and the scale dependent
NNLL+NNLO prediction normalized 
to this reference theoretical prediction. 
The scale dependence band of the perturbative calculation
is computed by varying $\mu_F$, $\mu_R$ and $Q$ as 
previously discussed in Sect.~\ref{sec:inclu}:
we vary $\mu_F$, $\mu_R$ and $Q$ in the range $m_Z/4\leq \{\mu_F,\mu_R,Q\}\leq m_Z$, 
with the constraints $0.5 \leq \{\mu_F/\mu_R,Q/\mu_R\} \leq 2$. 
We see that our perturbative calculation is consistent with the data within 
the uncertainties.
The scale variation bands at NLL+NLO and NNLL+NNLO accuracy overlap.
Moreover, in going from NLL+NLO to NNLL+NNLO accuracy
the perturbative uncertainty is reduced and the 
agreement between experimental data and theory prediction is improved.
The perturbative uncertainty at NNLL+NNLO accuracy is about $\pm 10$\% at the peak, it
decreases to about $\pm 4$\% at $q_T\sim 10$ GeV, 
and it increases again to about $\pm 10$\% at $q_T=40$ GeV.
The comparison between our theoretical prediction and the CMS and ATLAS data is qualitatively similar, 
the main difference being that, due to the larger data sample, 
the experimental errors in the ATLAS analysis are significantly smaller.

\begin{figure}[th]
\centering
\hspace*{-0.8cm}
\subfigure[]{
\includegraphics[width=3.53in]{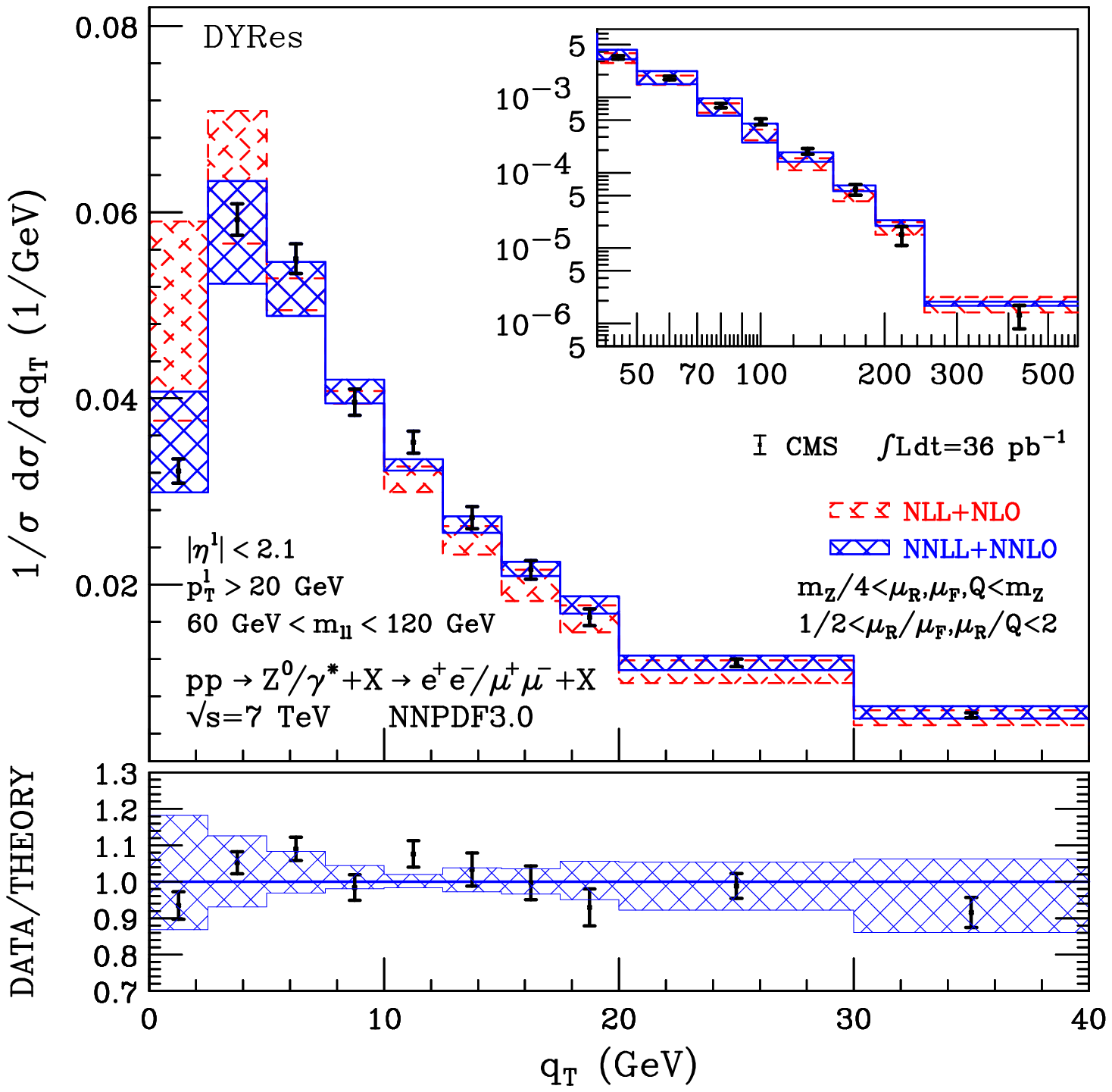}
}
\hspace*{.15cm}
\subfigure[]{
\includegraphics[width=3.28in]{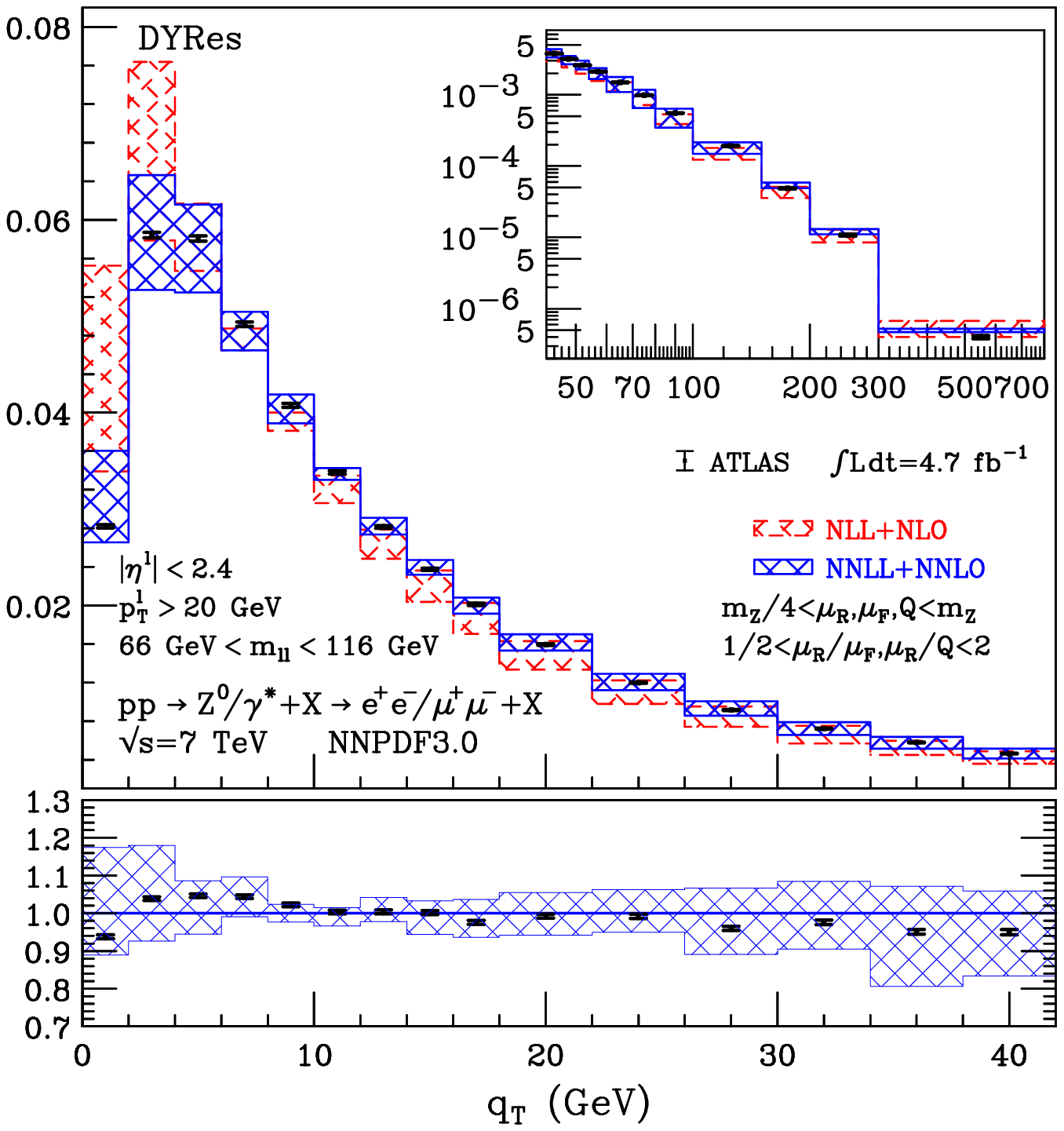}}
\caption{\label{fig:Zqt}
{\em Vector boson production at the LHC with lepton selection cuts.
The NLL+NLO (red) and NNLL+NNLO (blue) normalized  $q_T$ spectra for 
$Z/\gamma^*$ production are compared with the CMS data of 
Ref.~\cite{Chatrchyan:2011wt} (left panel)
and the ATLAS data of Ref.~\cite{Aad:2014xaa} (right panel).
The scale variation bands are obtained as described in the text.
The inset plot shows the ratio of the data and of the scale dependent 
NNLL+NNLO result with respect to the NNLL+NNLO result at central values of the
scales.
}}
\end{figure}

We add a comment on the large-$q_T$ region (see inset plots of
Fig.~\ref{fig:Zqt}), where the cross section is dominated
by the fixed-order contribution. 
For very large $q_T$, i.e. $q_T\gg m_Z$, the physical hard scale of 
the process is of the order of $q_T$ and not of the order of $m_Z$, 
and a sensible scale choice is $\mu_F \sim \mu_R \sim q_T$.
Therefore, it is not unexpected that our NNLL+NNLO calculations,
which use $\mu_F \sim \mu_R \sim m_Z/2$,
slightly overshoot the CMS and ATLAS 
data in the last few high-$q_T$ bins. The size of the QCD corrections 
evaluated with 
$\mu_F \sim \mu_R \sim q_T$
would be smaller. Moreover, in the extreme region $q_T\gg m_Z$
a resummation of enhanced large-$q_T$ perturbative terms is in principle
required~\cite{Berger:2001wr,Berger:2015nsa}.


\begin{figure}[!ht]
\centering
\includegraphics[width=0.8\textwidth]{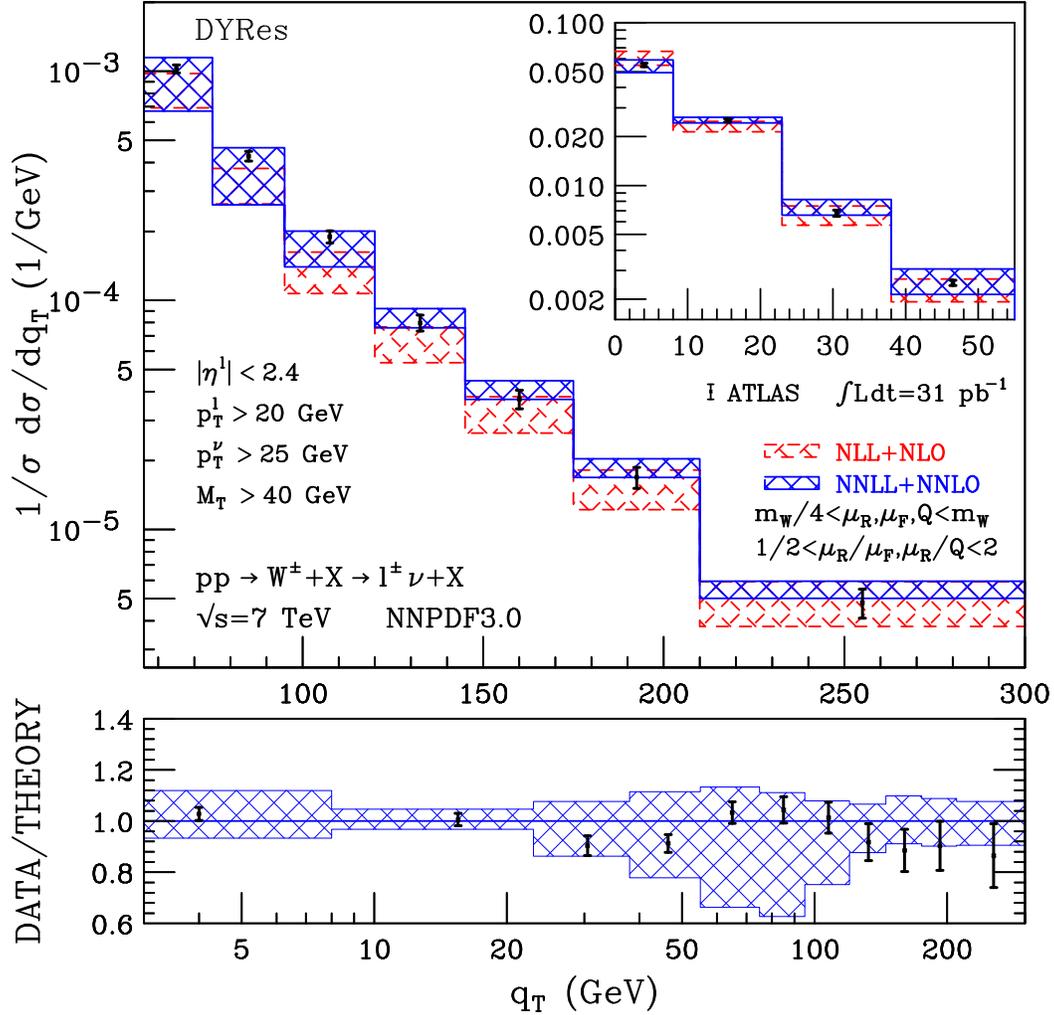}
\caption{{\em 
Vector boson production at the LHC with lepton selection cuts.
The NLL+NLO (red) and NNLL+NNLO (blue) normalized $q_T$ spectra for 
$W^\pm$ production are compared with the ATLAS data of 
Ref.~\cite{Aad:2011fp}.
The ratio in the lower panel and the scale variation bands are obtained as in
Fig.~\ref{fig:Zqt}.
}}
\label{fig:pTW}
\end{figure}

In Fig.~\ref{fig:pTW}
we consider the $q_T$ spectrum of $W^\pm$ bosons. We present a 
comparison of our resummed results with the $pp\to W\to l\nu$ data collected by the ATLAS 
Collaboration~\cite{Aad:2011fp}
with an integrated luminosity of 31~pb$^{-1}$ at 
$\sqrt{s}=7$~TeV. The fiducial region is defined as follows:
the charged lepton has transverse momentum 
$p_T^l>20$~GeV and pseudorapidity $|\eta^l|<2.4$, 
the missing transverse energy is $p_T^\nu>25$~GeV, and transverse mass 
$m_T=\sqrt{2p_T^l\, p_T^\nu(1-\cos(\phi^l-\phi^\nu))}$ is constrained in the
region $m_T>40$~GeV. 
We recall that, because of the presence of the neutrino in the final state, 
the  $q_T$ of the $W$ has to be
reconstructed through the transverse energy of the hadronic recoil, which has a poorer 
experimental resolution
than that of the lepton momentum.
In the small $q_T$ region, the bin sizes of the experimental data are rather 
large,
with only four bins in the region with $q_T < 55$~GeV. 
For this reason in Fig.~\ref{fig:pTW}
we focus on the large $q_T$ region 
55~GeV $< q_T < 300$~GeV, while the  small $q_T$ region is shown in 
the inset plot. 
The lower panel of Fig.~\ref{fig:pTW}, which covers the entire $q_T$ region
of the data, presents the ratio of both data and theoretical results with
respect to the reference theoretical result. This ratio and the scale variation
bands are computed exactly in the same manner as in the case of
Fig.~\ref{fig:Zqt}. 
Looking at the ratio plot in the lower panel,
we see that our NNLL+NNLO calculation describes the $W$ production data within the perturbative uncertainties.
The NNLL+NNLO perturbative uncertainty is about $\pm 8$\% at the peak, it
decreases to about $\pm 4$\% 
at $q_T\sim 15$~GeV, 
and it increases again to about $\pm 15$\% at $q_T=50$~GeV.


In Sect.~\ref{sec:inclu} and in the first part of this Section, we have examined
vector boson $q_T$ distributions (without and with the application of acceptance
cuts) and we have computed and studied the effects that are produced by the
all-order resummation of large logarithmically-enhanced terms at small values of
$q_T$. 
Our related calculations are performed at complete NNLL+NNLO 
(and NLL+NLO) accuracy. In the following part of this Section,
we consider other observables that are related to the $q_T$ distributions
but in which fixed values of $q_T$ are not directly measured. These observables are
inclusive over $q_T$ within certain $q_T$ ranges.
Since the bulk of the vector boson cross section is produced at small values of 
$q_T$, if the observable (indirectly) probes the {\em detailed} shape of the 
production cross section in the small-$q_T$ region, the observable itself can be
very sensitive to high-order radiative corrections and to the $q_T$ resummation
effects that we can explicitly compute. This 
reasoning
illustrates and justifies
the physical (and quantitative) relevance of $q_T$ resummation for other
$q_T$-related observables.
In the second
part of this Section we study the quantitative impact of $q_T$ resummation on some
observables.

At the formal level, our study of other observables implies that we are resumming 
high-order logarithmic corrections (in case they are present) 
that appear in the computation of those
observables. Strictly speaking, this resummation has to be performed on an
observable-dependent basis (see, e.g., Ref.~\cite{Banfi:2011dx}). Therefore,
our observable-independent treatment (based on transverse-momentum resummation)
cannot guarantee that we formally achieve exact NNLL+NNLO accuracy for all these
observables. Nonetheless we are able to correctly take into account all the
leading-logarithmic contributions,
all 
the complete (with and without logarithmic enhancement)
perturbative terms up to the
NNLO level\footnote{For observables that are inclusive over the region that
includes $q_T=0$, the NNLO accuracy is achieved through our detailed matching
procedure (see Sect.~\ref{sec:theory}) with the fixed-order calculation.}, 
and a substantial part of
subleading logarithmic terms beyond the NNLO accuracy. 

This statement about resummation is a consequence of the following discussion.
The observable-dependent logarithmic terms (in case they are present) are due to
multiple radiation of soft and collinear partons in the inclusive final-state:
these logarithmic corrections are computed by approximating the QCD scattering
amplitudes in the soft and collinear limits and, then, by integrating the
final-state QCD radiation over the corresponding phase space with appropriate
(observable-dependent) kinematical approximations. In our transverse-momentum
resummation procedure we correctly take into account the NNLL dynamics
(the behaviour of the QCD scattering
amplitudes in the soft and collinear limits) of soft and collinear radiation, and
we treat the phase space of the final-state QCD radiation with consistent 
kinematical approximations that are specific of the $q_T$ spectrum.
However, the observable-dependent kinematical approximations can only differ
beyond the leading-logarithmic level (to leading-logarithmic level, a 
strong-ordering approximation in the energy/angle of the emitted partons is
sufficient),  and these differences do not spoil the leading-logarithmic accuracy
of our resummed calculation. 
In this respect, it is important to remark the role of
the $q_T$ recoil (see Sect.~\ref{sec:theory} and Appendix~\ref{app})
on the kinematics of the produced (observed) lepton pair. We treat the 
$q_T$ recoil in a kinematically consistent way (though it necessarily involves non
logarithmic approximations that are {\em uniformly} of ${\cal O}(q_T/M)$
throughout the small-$q_T$ region), and such a treatment is necessary to 
correctly {\em correlate} the dynamical $q_T$ resummation effect with the ensuing 
$q_T$ dependence of the measured (computed) observable.

In summary, the application of our $q_T$ resummed calculations to the computation
of other observables is physically (and, thus, quantitatively) and formally
(as we have just discussed) justified. A detailed specification of the 
subleading-logarithmic accuracy of the  $q_T$ resummed calculation at the formal
(analytical) level requires (and deserves) observable-dependent investigations,
which can be performed in future studies.


\begin{figure}[!ht]
\centering
\includegraphics[width=0.9\textwidth]{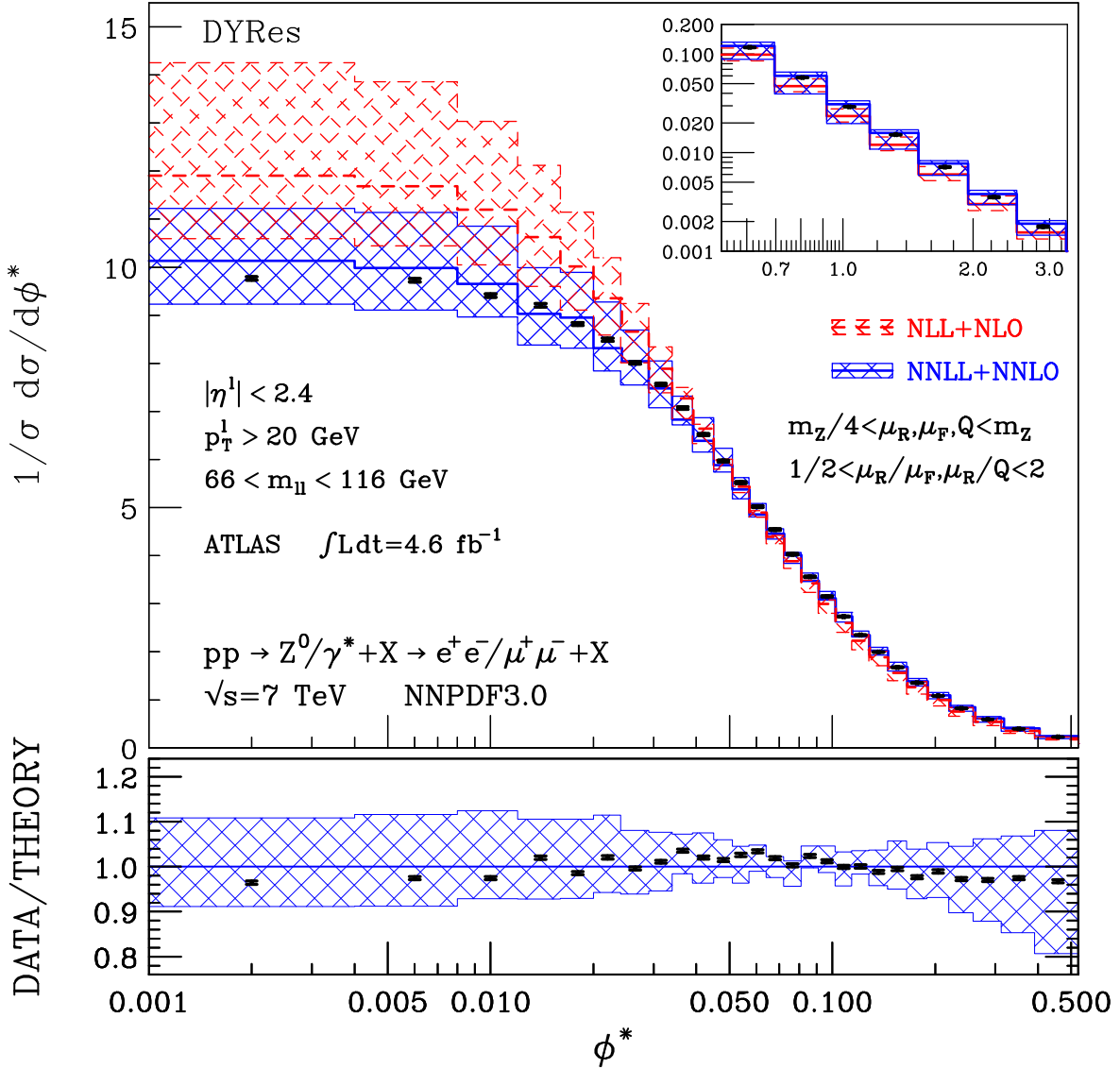}
\caption{{\em 
The NLL+NLO (red) and NNLL+NNLO (blue) normalized  $\phi^*$ distribution for
$Z/\gamma^*$ production at the LHC is compared with the ATLAS data of 
Ref.~\cite{Aad:2012wfa}. 
The NLL+NLO and NNLL+NNLO central results are computed at the scales
$\mu_R=\mu_F=Q=m_Z/2$.
The ratio in the lower panel and the scale variation 
bands are obtained as in
Fig.~\ref{fig:Zqt}.
}}
\label{fig:Wphi}
\end{figure}

\setcounter{footnote}{1}

Among other observables, we first consider
the measurement\footnote{An analogous measurement of the $\phi^*$ distribution
at the LHC  was reported by the LHCb Collaboration
\cite{Aaij:2012mda} with an integrated luminosity of $0.94~{\rm fb}^{-1}$.
In the small-$\phi^*$ region, the bin sizes of the LHCb measurement
are rather large (with respect to those of the ATLAS measurement
\cite{Aad:2012wfa}), with only two (four) bins in the region $\phi^* < 0.1$
($\phi^* < 0.2$).}
of the 
$\phi^*$ distribution from $pp\to ~Z/\gamma^*\to l^+l^-$ data 
at 
$\sqrt{s}=7$~TeV as reported by the ATLAS Collaboration~\cite{Aad:2012wfa}
with an integrated luminosity of 4.6 fb$^{-1}$.
The $\phi^*$ observable is defined as 
$\phi^*=\tan(\pi/2-\Delta\phi/2)\,\sin(\theta^*)$, where
$\Delta\phi$ is the azimuthal angle between the leptons and the 
angle $\theta^*$
is defined by 
$\cos\theta^*= \tanh((\eta^{l^+}-\eta^{l^-})/2)$ where $\eta^{l^+}$ 
($\eta^{l^-}$)
is the rapidity of the positively (negatively) charged 
lepton.
The cuts that define the fiducial region are those of 
the ATLAS analysis of the $q_T$ spectrum:  
66~GeV $< m_{ll} <$ 116~GeV, $p_T^l>20$~GeV and $|\eta^l|<2.4$.

The $\phi^*$ variable at small values of $\phi^*$ is correlated 
to $q_T$ and, therefore, it is strongly sensitive to $q_T$ resummation effects. 
A detailed discussion on the relation between $\phi^*$ and $q_T$ is presented in 
Ref.~\cite{Banfi:2011dx}, where the resummation of the $\ln \phi^*$ terms is
carried out in analytic form up to NNLL+${\cal O}(\as^2)$ 
accuracy\footnote{The analytical treatment of Ref.~\cite{Banfi:2011dx} does not
reach complete NNLO accuracy at small values of $\phi^*$ since the analogue of the
vector boson coefficient ${\cal H}_V^{(2)}$ in Eq.~(\ref{hexpan}) is not included
in the calculation. An approximated form of ${\cal H}_V^{(2)}$ is included in the
calculation of Ref.~\cite{Guzzi:2013aja}.}, 
and it turns out to be strictly related and very similar to 
$q_T$ resummation. Ensuing phenomenological studies are presented in 
Refs.~\cite{Banfi:2012du, Banfi:2011dm, Guzzi:2013aja}.

In Fig.~\ref{fig:Wphi} we report the ATLAS data of the $\phi^*$ distribution
(normalized to the measured cross section in the fiducial region) and the
comparison with the results of our resummed calculation.
The NLL+NLO and NNLL+NNLO central results are computed at the scales
$\mu_R=\mu_F=Q=m_Z/2$.
The scale variation bands at NLL+NLO (red) and NNLL+NNLO (blue) accuracy and the
reference NNLL+NNLO result for the ratio in the lower panel of Fig.~\ref{fig:Wphi}
are computed as in Figs.~\ref{fig:Zqt} and \ref{fig:pTW}.
We observe that the scale variation bands at the two subsequent orders overlap, 
and that the NNLL+NNLO perturbative uncertainty is substantially smaller 
than the NLL+NLO one. 
The NNLL+NNLO result is consistent with the data within the 
uncertainties
in both the small-$\phi^*$ and large-$\phi^*$ regions (the large-$\phi^*$ region
is shown in the inset plot).
The NNLL+NNLO perturbative uncertainty is about $\pm 10$\% for $\phi^*<0.01$, 
it decreases to about 
$\pm 5$\% at $\phi^*\sim 0.05$, 
and it increases again to about $\pm 10$\% at $\phi^*\sim 0.2$.


We add a comment on the results that we have shown in 
Figs.~\ref{fig:Zqt}--\ref{fig:Wphi}.
We recall that all the results presented in this Section are obtained 
in a purely perturbative framework. In Sect.~\ref{sec:inclu} we have discussed the
possible impact of the inclusion of a NP form factor, and we have seen 
(Fig.~\ref{fig:pdfs})
that NP effects should lead to a deformation of the perturbative result that
is well within the scale variation uncertainties
of the NNLL+NNLO calculation.
In Figs.~\ref{fig:Zqt}--\ref{fig:Wphi} we observe that all the resummed 
perturbative predictions are consistent 
with the data within our estimation of perturbative uncertainties.
Owing to the agreement between the theoretical NNLL+NNLO predictions and 
the experimental data
in the very small $q_T/\phi^*$ region,
we cannot draw any precise
quantitative conclusion about the definite size of NP effects
in the $Z/\gamma^*$, $W^\pm$ and $\phi^*$
distributions that we have considered.
We can only conclude that NP effects have to be small
in order not to spoil the agreement between the data and the corresponding
NNLL+NNLO results in Figs.~\ref{fig:Zqt}--\ref{fig:Wphi}.

\begin{figure}[!ht]
\centering
\includegraphics[width=0.9\textwidth]{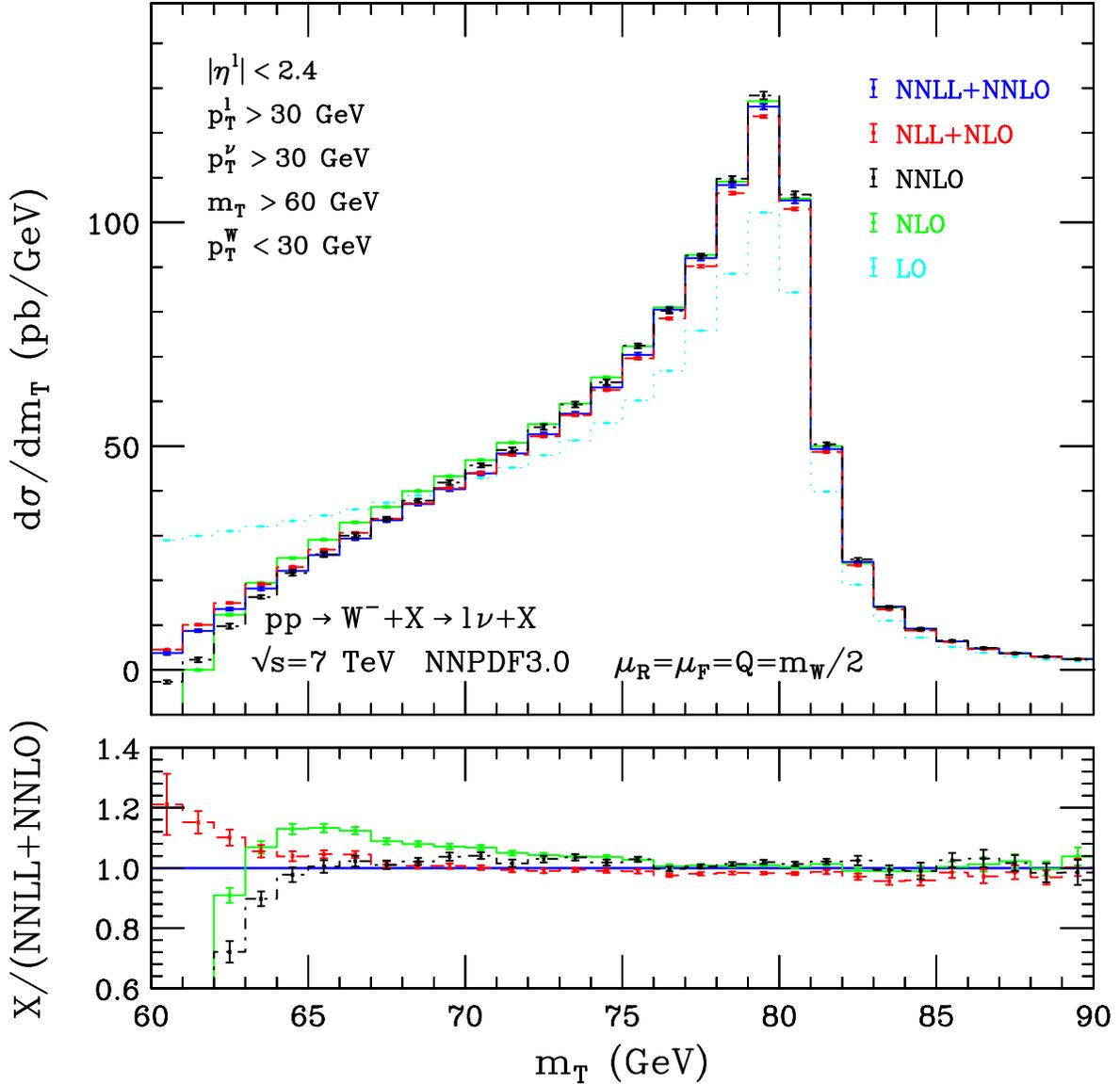}
\caption{{\em Effect of $q_T$ resummation on the transverse-mass ($m_T$)
distribution for $pp\to W^-\to l^-{\bar \nu}_l$ production 
at the LHC. Comparison of results of the fixed-order calculation 
at LO (cyan dotted), NLO (green solid) and NNLO (black dot-dashed) with the 
resummed calculation at NLL+NLO (red dashed) and 
NNLL+NNLO (blue solid) accuracy.
The lower panel shows the ratio between the various results 
(excluding the LO result) and the 
NNLL+NNLO result.
}}
\label{fig:Wtmass}
\end{figure}

\begin{figure}[th]
\centering
\hspace*{-0.3cm}
\subfigure[]{
\includegraphics[width=3.33in]{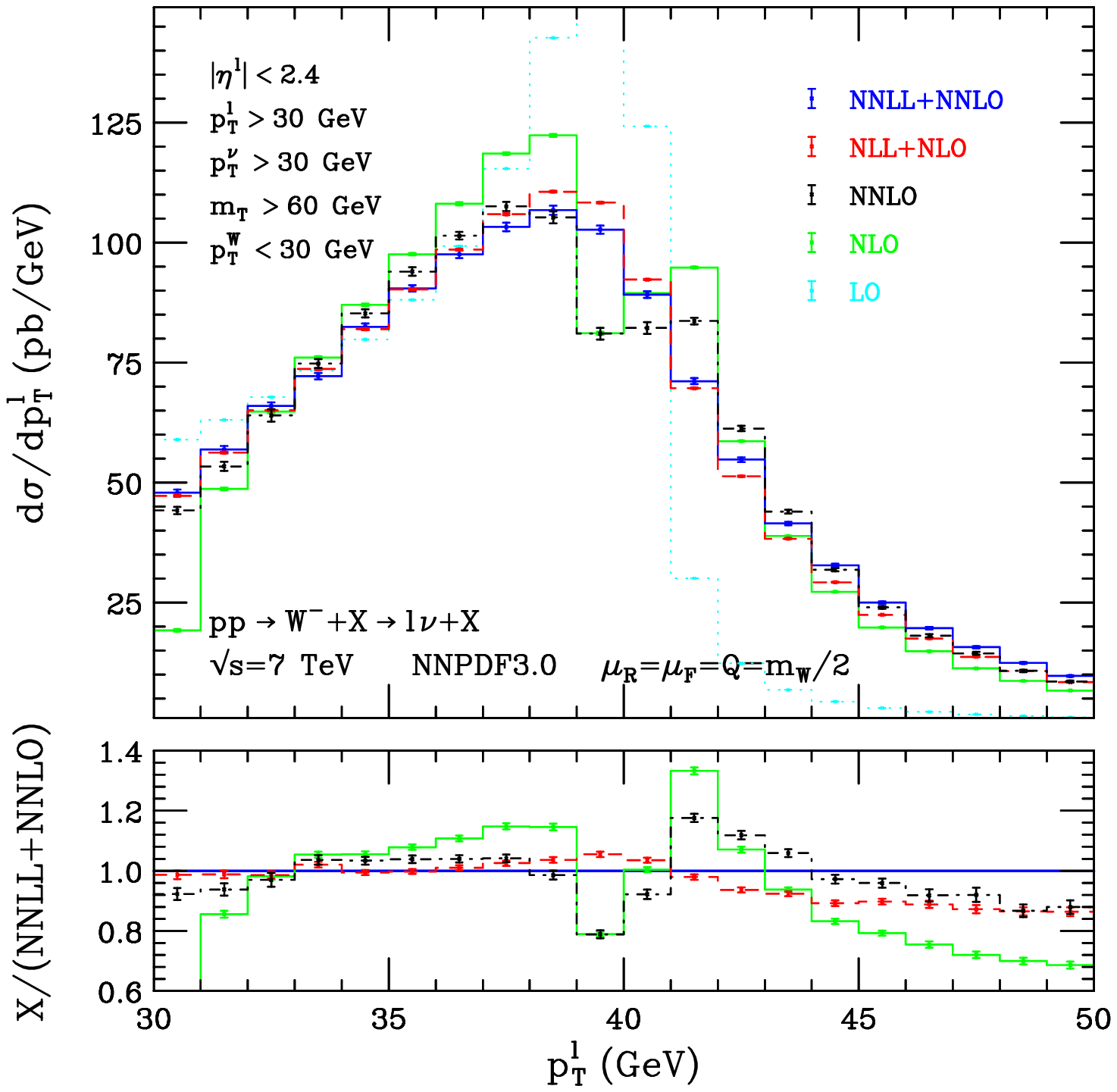}
}
\hspace*{.1cm}
\subfigure[]{
\includegraphics[width=3.33in]{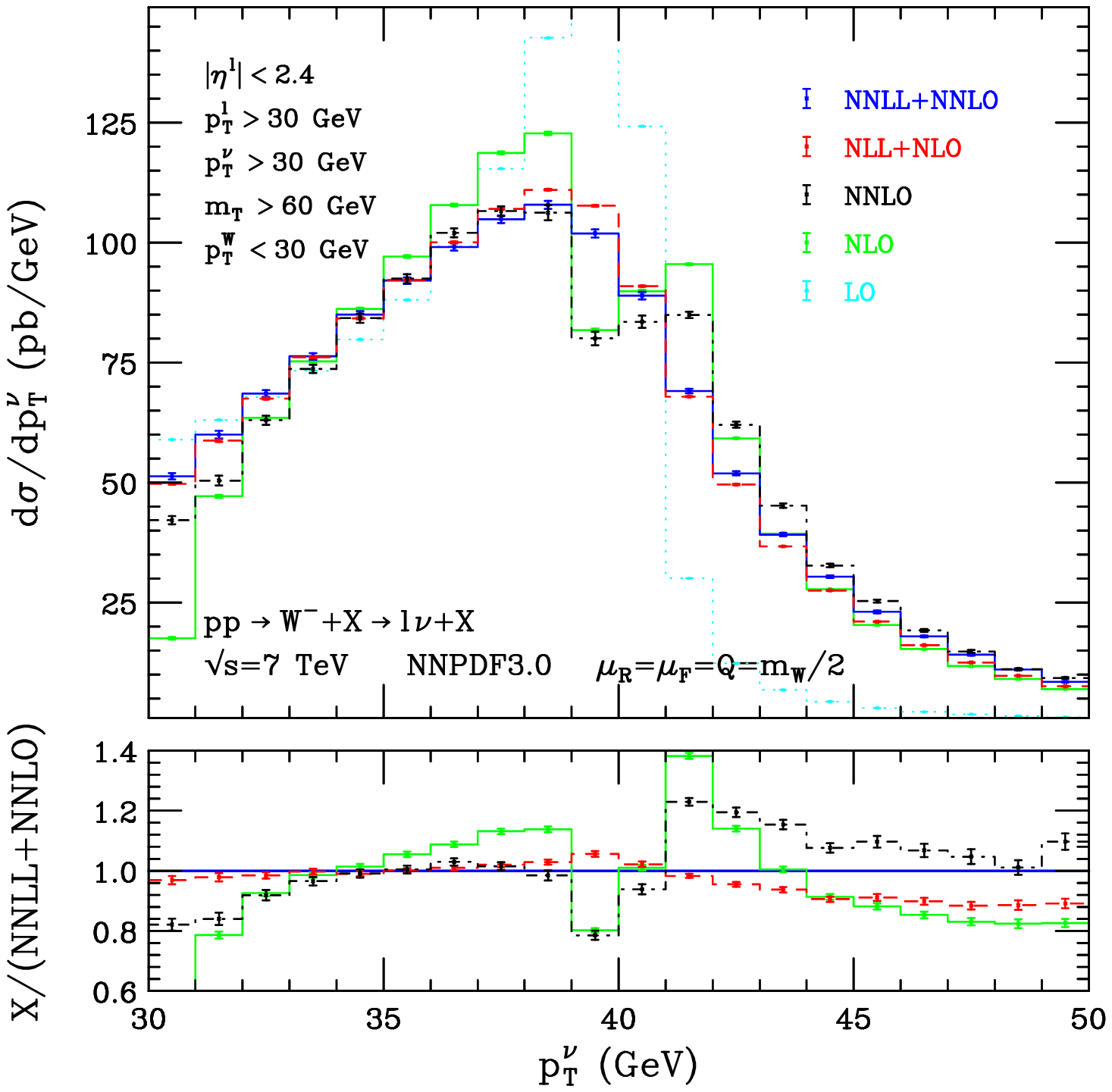}}
\caption{\label{fig:Wptlept}
{\em  Effect of $q_T$ resummation 
for $pp\to W^-\to l^-{\bar \nu}_l$ production 
at the LHC:
(a) lepton $p_T$ distribution and  (b) missing $p_T$ distribution.
The fixed-order and resummed results are denoted as in Fig.~\ref{fig:Wtmass}.
}}
\end{figure}


We conclude this Section by considering other observables.
We study the impact of $q_T$ resummation on the kinematical distributions 
that are relevant
for the measurement of the $W$ mass. 
We consider $pp\to W^-\to l^-{\bar \nu}_l$ with $\sqrt{s}=7$ TeV and 
we apply the following selection cuts: the charged lepton has
transverse momentum 
$p_T^l>30$~GeV and rapidity $|\eta^l|<2.4$, 
the missing transverse 
momentum
is $p_T^\nu>30$~GeV, and the transverse mass $m_T$
has $m_T>60$~GeV. We also apply a cut, $p_T^W<30$~GeV, on the transverse momentum 
$p_T^W$ of the $W$ boson (lepton pair). 
The results of our calculation of the $m_T$ distribution and of the lepton momentum
distributions are presented in Fig.~\ref{fig:Wtmass} and \ref{fig:Wptlept},
respectively. The reference scale choice of the calculation is
$\mu_F=\mu_R=Q=m_W/2$.
In both figures we present the results of the fixed-order calculation 
at LO (cyan dotted),
NLO (green solid) and NNLO (black dot-dashed) accuracy  and we compare them with the
results of the $q_T$ resummed calculation at NLL+NLO (red dashed) and 
NNLL+NNLO (blue solid) accuracy.
The lower panels show the ratio between the various results and the 
NNLL+NNLO result (the ratio LO/(NNLL+NNLO) is not reported in the lower panels).


The $m_T$ distribution in the range $m_T < 90$~GeV is presented in 
Fig.~\ref{fig:Wtmass}. We can consider two regions: the large-$m_T$ region,
around $m_T \sim m_W$ (we recall that we use $m_W = 80.385$~GeV), and the 
small-$m_T$ region.
In the large-$m_T$ region, $m_T\gtap 70$~GeV, we see that the perturbative 
prediction is extremely stable against radiative corrections, and the stability
is present both in going from NLO to NNLO accuracy and with inclusion of resummation.
This is a consequence of the well known fact that the transverse mass is 
weakly sensitive to the transverse momentum of the $W$ boson. 
Formally, the $m_T$ distribution has 
no logarithmic
corrections of the type $\ln(|m_T - m_W|/m_W)$, and our $q_T$ resummed calculation
does not spoil the stability of the fixed-order expansion.
On the contrary, in the small-$m_T$ region, we observe that
the fixed-order predictions become unreliable. The LO distribution is large at
$m_T = 60$~GeV, and 
both the NLO and NNLO distributions become negative
at $m_T\sim 60$ GeV.  This (mis-)behaviour is due to the fact that the constraints
$p_T^l>30$~GeV and $p_T^\nu>30$~GeV produce an unphysical boundary (and a stepwise
behaviour) of the $m_T$ distribution at $m_T=m_{T {\rm step}}= 60$~GeV in the LO
calculation. The boundary is due to the LO kinematics 
$\bpt^l+\bpt^\nu= \bqt =0$, and it disappears at higher orders since $q_T \neq 0$.
The LO  boundary induces (integrable) logarithmic singularities of the type
$\ln(1- m_{T {\rm step}}/m_T)^2$ at NLO and beyond \cite{Catani:1997xc}.
These logarithmic terms are resummed to all order by $q_T$ resummation, and the 
singularities are absent 
in the resummed prediction \cite{Catani:1997xc}, 
which is well behaved at the LO boundary 
$m_T=m_{T {\rm step}}$. We also note that the differences between
the NLL+NLO and 
NNLL+NNLO results are small at $m_T \sim 60$~GeV.

In Figs.~\ref{fig:Wptlept}~(a) and (b) we present the $p_T^l$ and $p_T^\nu$ 
distributions, respectively.
In the limit in which the $W$ boson is produced
on shell, these distributions have an LO kinematical boundary at $m_W/2$.
The finite width of the $W$ boson (partially)
smears this effect: at LO both the $p_T^l$ and $p_T^\nu$ distributions are strongly peaked at $m_W/2$ (Jacobian peak)
and quickly drop for $p_T\gtap m_W/2$. 
The almost stepwise behaviour of the LO distribution produces 
large radiative corrections at NLO and beyond
(in the limit in which the $W$ boson is produced
on shell, these large corrections would be 
integrable logarithmic singularities at each perturbative order 
\cite{Catani:1997xc}).
The NLO and NNLO distributions indeed display an unphysical peak at 
$p_T\sim 42$~GeV,
which is an artifact of such large corrections (singularities in the on-shell 
limit).
The resummed predictions at NLL+NLO and NNLL+NNLO accuracy are free of
such instabilities and display a smooth {\it shoulder} behaviour around the 
LO boundary for on-shell production.
The perturbative instabilities of the fixed-order calculation at small values
of $p_T$ ($p_T^l\sim 30$~GeV and $p_T^\nu \sim 30$~GeV) are analogous to those
that we have previously discussed in the case of the $m_T$ distribution in the
region $m_T\sim 60$~GeV 
(see Fig.~\ref{fig:Wtmass}). In the case of the $p_T$ distributions, it is the
constraint $m_T > 60$~GeV that produces the LO boundaries at 
$p_T^l=p_T^\nu =30$~GeV, an LO stepwise behaviour and ensuing  
instabilities at each subsequent perturbative order.
The resummed calculation is perturbatively stable in the small-$p_T$ region,
and the differences between
the NLL+NLO and 
NNLL+NNLO results are small throughout the entire region with $p_T\ltap 45$~GeV.
In the large-$p_T$ region ($p_T\gtap 45$~GeV)
both the $p_T^l$ and $p_T^\nu$ distributions display radiative corrections that are relatively large.
This is not unexpected since in this region of transverse momenta the
NLO calculation is essentially
the first perturbative order at which both the $p_T^l$ and the $p_T^\nu$ 
distributions are non vanishing (in the on-shell limit, the ${\cal O}(\as)$ and 
${\cal O}(\as^2)$ result would be an LO and an NLO prediction, respectively).

\section{Summary}
\label{sec:summary}

In this paper we have considered the transverse-momentum ($q_T$) distribution of 
DY high-mass lepton pairs produced, via $Z/\gamma^*$ and $W$ bosons decay,
in hadronic collisions.
We have presented a perturbative QCD study based on 
transverse-momentum resummation
up to NNLL accuracy.
We have combined small-$q_T$ resummation with the known ${\cal O}(\as^2)$ 
fixed-order result at small, intermediate and large values of $q_T$.

We have followed the resummation formalism developed 
in Refs.~\cite{Catani:2000vq,Bozzi:2005wk,Bozzi:2007pn}
to implement transverse-momentum resummation and the matching with the result at 
${\cal O}(\as^2)$. In particular, our calculation includes the complete NNLO
contributions at small values of $q_T$ (i.e., in any regions that include 
$q_T=0$) and it exactly reproduces the complete NNLO total cross section
after integration over $q_T$. 
This leads to theoretical predictions with a controllable and uniform 
perturbative accuracy over the region
from small up to large values of $q_T$. At large values of $q_T$, the 
predictivity of small-$q_T$ resummation is superseded by that of the customary
fixed-order expansion, and our resummed calculation can be smoothly joined
onto the ${\cal O}(\as^2)$ calculation. The resummed calculation can be
systematically expanded at various orders of logarithmic accuracy
(e.g., NLL+NLO and NNLL+NNLO accuracy), and its theoretical uncertainties
due to uncalculated higher-order QCD corrections can be studied by comparing the
results at two subsequent orders and by performing 
systematic studies on
factorization, renormalization and resummation scale dependence.
We have performed such a study for the case of vector boson production 
at LHC energies, 
and we have briefly illustrated the uncertainties due to 
parton densities and the possible impact of non-perturbative effects.

In the present paper we have extended
the resummed calculation presented in Ref.~\cite{Bozzi:2010xn} for
$Z/\gamma^*$ production 
by considering also $W^\pm$ production and by including
the leptonic decay of the vector boson with the corresponding spin correlations, 
the finite-width effects
and the full dependence on the final-state leptonic variables.
We have compared  our resummed results for $Z/\gamma^*$  and $W$ production 
with some of the available data of  
the ATLAS and CMS experiments at the LHC, 
applying the same kinematical cuts on final state leptons
that are considered in the experimental analyses. 
We find that the data are well described by our predictions
within the perturbative uncertainties.
We have also considered the impact of transverse-momentum resummation 
on observables, which are different from the vector boson $q_T$, that
depend on the lepton kinematical variables. In particular,
we have studied
the $\phi^*$ distribution
in $Z/\gamma^*$ production and the leptonic transverse-momentum, 
the missing 
transverse-momentum
and the transverse-mass distributions in $W$ production.

Our calculation is implemented in the parton-level Monte Carlo numerical code {\tt DYRes} which
allows the user to apply arbitrary kinematical cuts on the vector boson 
and the final-state leptons,  
and to compute the corresponding relevant distributions in the form of bin histograms.
These features 
make our program a useful tool for DY studies at the Tevatron and the LHC.
A version of the  {\tt DYRes} code is publicly available.

The production and decay mechanisms of the vector boson are dynamically 
correlated by the non vanishing spin of the vector boson.
The inclusion of the lepton decay (with the spin correlations and the 
full dependence on the kinematical variables of the two leptons)
in the resummed calculation requires a general theoretical discussion on the 
$q_T$ recoil due to the transverse momentum of the vector boson.
This discussion is not limited to the specific case of vector boson production.
We have presented a general and explicit procedure to treat the $q_T$ recoil.
The procedure is directly applicable to 
$q_T$ resummed calculations for production processes of generic
high-mass systems in hadron collisions.

\paragraph{Acknowledgements.} 
We would like to thank Stefano Camarda, Luca Perrozzi and Jan Stark for useful discussions.
This research was supported in part by the Swiss National Science Foundation 
(SNF) under contracts CRSII2-141847, 200021-156585 and by the Research 
Executive Agency (REA) of the 
European Union under the Grant Agreement number PITN-GA-2012-316704 
({\it Higgstools}).

\appendix

\section{Appendix: Lepton angular distribution and $q_T$ recoil in
transverse-momentum resummation}
\label{app}

This Appendix is devoted to the $q_T$-recoil issue that we have introduced and
illustrated in Sect.~\ref{sec:theory} 
(see Eqs.~(\ref{Frecoil})--(\ref{Frecoilform}) and
accompanying comments).
To our knowledge the issue has not received much attention in the previous
literature on transverse-momentum resummation. We present a
detailed discussion of
the issue and a general, explicit and consistent procedure to implement the 
$q_T$ recoil in transverse-momentum resummation. Our procedure explicitly 
exhibits the degree of freedom involved in the implementation of the  
$q_T$ recoil and, moreover, it gives an explicit formal parametrization of the
ensuing ambiguities. The procedure is straightforwardly applicable to implement
the $q_T$ recoil (and, possibly, estimate related uncertainties) in calculations
based on transverse-momentum resummation.

The $q_T$-recoil issue is not specific of the lepton angular distribution for
vector boson decay, but it regards transverse-momentum resummation for generic
production processes. For simplicity of presentation, in the following we consider
in detail vector boson production and the DY process. Then we discuss the
generalization to generic processes.

We begin our discussion by considering the computation of the DY
multidifferential cross section in Eq.~(\ref{fact}) at the LO in perturbative QCD.
At this order the hadronic cross section (and the corresponding partonic 
cross section) is directly and exactly (i.e., with no small-$q_T$ approximation)
proportional to the Born level angular distribution 
$d{\hat \sigma}^{(0)}/d{\bomega}$ in Eq.~(\ref{resum}).
We have
\begin{equation}
\label{distlo}
\frac{d\sigma_{h_1h_2\to l_3l_4}}{d^2{\bqt}\,d{M^2}\,dy\,d{\bomega}} 
(\bqt,M,y,s,{\bomega}) \;
\propto \; \left[ \frac{{d{\hat \sigma}^{(0)}_{a_1a_2\to l_3l_4}}}{d{\bomega}}
\right]_{\rm LO} \;\;
\delta^{(2)}(\bqt) \;\;.
\end{equation}

For the purpose of our general discussion of $q_T$ recoil, we write the lepton
angular distribution in the following form:
\begin{eqnarray}
d{\bomega} \;\;\;
\frac{{d{\hat \sigma}^{(0)}_{a_1a_2\to l_3l_4}}}{d{\bomega}}(k_1,k_2;p_3,p_4)
& \propto &
\frac{1}{M^2} \int d^4p_3 \;d^4p_4\;\delta_+(p_3^2) \;\delta_+(p_4^2)
\;\delta^{(4)}(q-p_3-p_4) \nn \\
\label{distall}
& \times & | M^{(0)}_{a_1a_2\to l_3l_4}(k_1,k_2;p_3,p_4) |^2 \;\;.
\end{eqnarray}
Note that, following the general notation of Eq.~(\ref{fact}), we have not
specified the actual definition of the angular variables $\bomega$, and we have
written the left-hand side of Eq.~(\ref{distall}) in a Lorentz invariant form.
The relation (\ref{distall}) is written in the form of a proportionality relation:
the additional proportionality factors that are not explicitly denoted in the 
right-hand side are not relevant for our following discussion 
(in particular, they are independent of the lepton momenta $\{p_3,p_4\}$ and,
thus, of $\bomega$). The factor
$| M^{(0)}_{a_1a_2\to l_3l_4}|^2$ is the square of the Born level scattering
amplitude $M^{(0)}_{a_1a_2\to l_3l_4}$  for the partonic process
\begin{equation}
\label{parpro}
a_1(k_1) + a_2(k_2)  \to \;\ell_3(p_3) + \ell_4(p_4) \;\;,
\end{equation}
where $k_i$ $(i=1,2)$ is the momentum of the colliding parton $a_i$ from the 
initial-state hadron $h_i(P_i)$ (see Eq.~(\ref{first})),
with the kinematics 
\begin{equation}
\label{kinem}
k_1+k_2 = q \;\;, \quad \;\; k_i^2 = 0 \;\;(i=1,2) \;\;.
\end{equation}
In our specific case of vector boson production, the Born level partonic process
is the $q{\bar q}$ annihilation process $q_f\bar q_{f'}\to V \to l_3l_4$
(i.e., $\{ a_1,a_2 \} = \{ q_f, {\bar q}_{f'} \}$).
All the other factors in the right-hand side of the relation (\ref{distall})
are related to the kinematical phase space of the final-state leptons and, in
particular,  they enforce the kinematical constraint 
$q= p_3 + p_4$.

The LO calculation of the cross section kinematically relates the parton and
hadron momenta $k_i$ and $P_i$. In particular, at the LO we have $\bqt =0$ and,
specifically, the LO value $k_{i \,({\rm LO})}$ of the parton momentum 
is\footnote{The kinematical variables $x_i$ in Eqs.~(\ref{lok12}) and
(\ref{xidef}) and the kinematical variable $z_1$ in Eq.~(\ref{z1res}) should not be
confused with the integration variables $x_i$ and $z_1$ 
used in Sect.~\ref{sec:theory} (we use the same symbols for both set of
variables).}
\beq
\label{lok12}
k_{1 \,({\rm LO})}^\mu = x_1 P_{1}^\mu\;\;,\quad \quad
k_{1 \,({\rm LO})}^\mu = x_2 P_{2}^\mu \;\;, 
\eeq
with
\beq
\label{xidef}
x_1 = \frac{M \;e^{\,+y}}{\sqrt s}\;\;,\quad \quad
x_2 = \frac{M \;e^{\,-y}}{\sqrt s} \;\;,
\eeq
where (see Sect.~\ref{sec:theory}) $M$ and $y$ are the invariant mass and the
rapidity of the lepton pair and ${\sqrt s}$ is the hadronic centre--of--mass
energy.
Inserting the LO expression (\ref{lok12}) of the parton momenta $k_i$
in Eq.~(\ref{distall}), the LO lepton angular distribution
$\left[ d{\hat \sigma}^{(0)}_{a_1a_2\to l_3l_4}/d{\bomega} \right]_{\rm LO}$ 
in Eq.~(\ref{distlo}) is uniquely specified.

Higher-order perturbative contributions produce logarithmically-enhanced
(`singular') terms at small $q_T$ that can be resummed to all orders, leading to
the resummation factor ${\hat W}$ in Eq.~(\ref{resum}).
These logarithmic terms are due to multiple radiation of soft and collinear
partons, and this soft and collinear radiation is {\em factorized} 
\cite{Catani:2013tia} with respect to the Born level amplitude 
$M^{(0)}_{a_1a_2\to l_3l_4}$ of Eq.~(\ref{distall}).
As a consequence,
after $\qt$ resummation the angular distribution of the decaying leptons
is still given by the Born level function $d{\hat \sigma}^{(0)}/d{\bomega}$
in Eq.~(\ref{distall}), and this function thus appears as a multiplicative
factor in front of the resummation factor ${\hat W}$ of the resummed component
of the vector boson $q_T$ cross section (see Eq.~(\ref{resum})).
{\em Strictly speaking} \cite{Catani:2013tia}, in the limit $q_T~\ll~M$ that is
relevant for resummation, the angular distribution can be expressed
in terms of the LO 
distribution
$\left[ d{\hat \sigma}^{(0)}/d{\bomega} \right]_{\rm LO}$ in Eq.~(\ref{distlo}),
namely the expression (\ref{distall}) with the LO
kinematics of Eqs.~(\ref{lok12}) and (\ref{xidef}), which in particular
has $\bqt = 0$. Indeed, after soft/collinear factorization and resummation,
any residual dynamical effect on the process in Eq.~(\ref{parpro})
(and on $M^{(0)}$ and $d{\hat \sigma}^{(0)}/d{\bomega}$) is due to hard-parton
radiation. Hard radiation produces ${\cal O}(q_T/M)$ corrections that lead to
non-singular contributions if $q_T~\ll~M$: these corrections can be formally
approximated by their limiting behaviour as $q_T \to 0$ and, thus, neglected 
in the computation of the resummed component 
(see Eqs.~(\ref{resum}) and (\ref{resumydiff}))
and included in the finite component (see Eq.~(\ref{resplusfin})).

Neglecting these ${\cal O}(q_T/M)$ corrections is a perfectly suitable procedure
for the resummed calculation of the vector boson $q_T$ cross section 
(see Eq.~(\ref{resumydiff})).
However, performing the resummation at fixed lepton momenta,
the momentum of the vector boson must be fully specified by the lepton momenta
and, in particular, 
$ \bqt = {\bpt}_{3} + {\bpt}_4$ is not vanishing.
The resummation factor $\hat W$ (see Eq.~(\ref{resum})) produces a smearing of the
LO distribution $\delta^{(2)}(\bqt)$ of Eq.~(\ref{distlo}) and finite values 
of $q_T$: to avoid unphysical results (e.g., events with $\bqt \neq 0$ and
${\bpt}_{3} + {\bpt}_4 = 0$) the factor $d{\hat \sigma}^{(0)}/d{\bomega}$
in Eq.~(\ref{resum}) cannot be the LO angular distribution 
$\left[ d{\hat \sigma}^{(0)}/d{\bomega} \right]_{\rm LO}$ (which has 
${\bpt}_{3} + {\bpt}_4 = 0$) in Eq.~(\ref{distlo}).
In other words, the non-vanishing value of $q_T$ has to be distributed between
the two lepton momenta and this leads to the $q_T$-recoil issue that we have
illustrated in Sect.~\ref{sec:theory} 
(see Eqs.~(\ref{Frecoil})--(\ref{Frecoilform}) and
accompanying comments). The resummed calculation requires the specification of 
a $q_T$-recoil 
prescription that has to be consistent (and physically sensible),
although this can be done in many (infinitely many) different ways.

Actual resummed calculations performed in the literature do not mention the 
$q_T$-recoil issue. The calculations of 
Refs.~\cite{Balazs:1995nz,Balazs:1997xd,Ellis:1997sc}
directly refer to the use of the Collins--Soper (CS) rest frame 
\cite{Collins:1977iv}. The procedure to compute the factor 
$d{\hat \sigma}^{(0)}/d{\bomega}$ in Eq.~(\ref{resum}) is as follows.
The lepton angular variables $\bomega$ are specified to be the polar and
azimuthal angles $\{ \theta^\prime_{CS} , \phi^\prime_{CS} \}$ of one of the
leptons in the CS rest frame. The LO distribution 
$\left[ d{\hat \sigma}^{(0)}/d{\bomega} \right]_{\rm LO}$ in Eq.~(\ref{distlo})
is then expressed in terms of $\{ \theta^\prime_{CS} , \phi^\prime_{CS} \}$
(since the LO distribution has $q_T =0$, in this case 
$\{ \theta^\prime_{CS} , \phi^\prime_{CS} \}$ exactly coincide with the lepton
scattering angles in the centre--of--mass frame of the LO colliding parton momenta in
Eq.~(\ref{lok12})) and this leads to an unambiguously defined angular function
$F_{q_f {\bar q}_{f^\prime} \to l_3l_4}(\theta^\prime_{CS} , \phi^\prime_{CS})$
(this function is actually independent of $\phi^\prime_{CS}$) that is used to
define (see Eq.~(\ref{Frecoil})) the angular distribution 
$d{\hat \sigma}^{(0)}/d{\bomega}$ of the resummed component of the cross section
(see Eq.~(\ref{resum})). This is a perfectly defined and consistent procedure, but
it hides the actual implementation of ${\cal O}(q_T/M)$ corrections through
an implicit prescription for the $q_T$ recoil: the definition of the 
CS rest frame is $q_T$ dependent and a $q_T$ dependence is introduced by
identifying/equating the angles $\{ \theta^\prime_{CS} , \phi^\prime_{CS} \}$
of the LO and resummed calculations (additional comment on this are presented in
a paragraph after Eq.~(\ref{k12asym})).

Here we explicitly present a consistent $q_T$-recoil procedure and an entire class
of $q_T$-recoil prescriptions. Our viewpoint is as follows: the non-vanishing
value of $q_T$ of dynamical origin that is produced by resummation leads to a 
$q_T$-recoil that can be 
`kinematically absorbed'~\footnote{The $q_T$ recoil issue does not arise in the 
context of transverse-momentum $(k_T)$ factorization 
\cite{Catani:1990xk}
for high-energy (small-$x$) hard-scattering processes. In this formulation the 
$q_T$ recoil is dynamically (and uniquely) embedded in the factorization formula.
The parton densities of the colliding hadrons are $k_T$ dependent and the
hard-scattering colliding partons have ensuing non-vanishing transverse momenta 
$\bkit$ that enter as integration variables in the factorization formula.}
by the momenta 
$k_1$ and $k_2$ of the colliding partons of the underlying hard-scattering 
process (see Eq.~(\ref{parpro})). As specified below, there are infinitely-many 
ways of implementing this kinematical recoil on the colliding partons 
in a consistent manner (i.e., without modifying the logarithmically-enhanced
perturbative terms at small $\qt$): they differ by corrections that are 
of ${\cal O}(\qt/M)$
order-by-order in the perturbative expansion 
(after having matched the resummed calculation with the complete N$^k$LO 
calculation, 
as in Eqs.~(\ref{resplusfin}) and (\ref{fincomp}),
these corrections start to contribute at the N$^{k+1}$LO level).

According to our procedure, the lepton angular distribution 
$d{\hat \sigma}^{(0)}/d{\bomega}$ to be used in the resummed calculation
(see Eq.~(\ref{resum})) is exactly given by the expression in 
Eq.~(\ref{distall}). The phase space factor in the right-hand side of 
Eq.~(\ref{distall}) is directly given in terms of the physical (measured) lepton
momenta $p_3$ and $p_4$ (with $p_3 +p_4=q$). The momentum $k_1$
(then, $k_2=q-k_1$) to be used to compute the Born level scattering amplitude in 
Eq.~(\ref{distall}) is given by the following parametrization:
\beq
\label{k1res}
k_1^\mu = z_1 \;\frac{M^2}{2q\cdot P_1} \;P_1^\mu + k_{1 T}^\mu +
\frac{\bkonet^2 }{z_1} \;
\frac{q\cdot P_1}{M^2 P_1 \cdot P_2} \;P_2^\mu \;\;, \quad \quad 
(k_{1 T}^\mu k_{1 T \mu} =
-\bkonet^2) \;\;,
\eeq 
where
\beq
\label{z1res}
z_1 = \frac{ M^2 + 2 {\bqt} \cdot \bkonet + 
{\sqrt {(M^2 + 2 {\bqt} \cdot \bkonet)^2 
- 4 M_T^2 \,\bkonet^2}} }{2 M^2} \;\;, \quad
(M_T^2 \equiv M^2 +q_T^2) \;\;,
\eeq
and $k_{1 T}^\mu$  is a two-dimensional vector that is 
transverse to both  $P_1^\mu$ and $P_2^\mu$ (i.e., $\bkonet$ lies in the 
$\bqt$ plane) and that fulfils the
following constraints:
\beq
\label{ktsmooth}
\bkonet \to 0 \quad  {\rm if}  \quad  \bqt \to 0 \;\;,
\eeq
\beq
\label{ktmax}
M^2 + 2 \bqt \cdot \bkonet > 2 \,M_T \;|\bkonet|
\;\;.
\eeq

We note that, following the definition in Eqs.~(\ref{k1res}) and (\ref{z1res}),
$k_1^\mu$ and $k_2^\mu$ are well defined `physical' parton momenta: they fulfil the
kinematics in Eq.~(\ref{kinem}) and they have positive definite energies,
$k_1^0 > 0$ and $k_2^0 > 0$
(the constraint in Eq.~(\ref{ktmax}) guarantees that the four-momentum
$k_{1}^\mu$ has positive definite energy and, then, $k_{2}^0 > 0$ follows
from $q^0 > 0$). Therefore the scattering amplitude 
$M^{(0)}_{a_1a_2\to l_3l_4}(k_1,k_2;p_3,p_4)$ in Eq.~(\ref{distall})
is well defined and unambiguously computable. Moreover,
due to Eq.~(\ref{ktsmooth}), the parton momentum $k_1$ in Eq.~(\ref{k1res})
coincides with its LO expression (\ref{lok12}) if $q_T = 0$.
We also note that $k_1^\mu$ is invariant under longitudinal boosts
of the hadronic centre--of--mass frame, provided $k_{1 T}^\mu$ is boost invariant.

At fixed values of $q^{\mu}, P_1^\mu$ and $P_2^\mu$, 
Eqs.~(\ref{k1res})--(\ref{ktmax}) give the most general expression 
of $k_1$ that respects the Born level kinematics in Eqs.~(\ref{parpro}) and
(\ref{kinem}) and the LO kinematics in Eqs.~(\ref{lok12}) and (\ref{xidef}).
This expression is parametrized by the arbitrary
(though constrained) 
transverse-momentum vector $\bkonet$. By choosing different values of 
$\bkonet$, we can obtain an entire class of consistent $q_T$-recoil
prescriptions. For example, two `obvious' possible choices are as follows:

A) set $\bkonet= \bqt/2$ (and thus $\bktwot= \bqt/2$):

\noindent from Eq.~(\ref{z1res}) we obtain
\beq
\label{csz1}
z_1 = \frac{M_T + M}{M_T} \;\;\frac{q\cdot P_1 \;\;q\cdot P_2}{M^2 \;P_1\cdot P_2}
\;\;,
\eeq
and we have
\beeq
\label{csk1}
k_1^\mu &=& \frac{M_T + M}{2 M_T} \;\frac{q\cdot P_2}{P_1\cdot P_2} \;P_1^\mu 
+ \frac{1}{2} \; q_{T}^\mu +
\frac{M_T - M}{2 M_T} \;\frac{q\cdot P_1}{P_1\cdot P_2} \;P_2^\mu \;\;, \\
\label{csk2}
k_2^\mu &=& 
\frac{M_T - M}{2 M_T} \;\frac{q\cdot P_2}{P_1\cdot P_2} \;P_1^\mu 
+ \frac{1}{2} \;  q_{T}^\mu +
\frac{M_T + M}{2 M_T} \;\frac{q\cdot P_1}{P_1\cdot P_2} \;P_2^\mu \;\;,
\eeeq

B) set $\bkonet= 0$ (and thus $\bktwot= \bqt$):

\noindent from Eq.~(\ref{z1res}) we obtain $z_1=1$ and we have
\beq
\label{k12asym}
k_1^\mu = \frac{M^2}{2 q\cdot P_1} \;P_1^\mu \;\;, \quad 
k_2^\mu = q^\mu - \frac{M^2}{2 q\cdot P_1} \;P_1^\mu \;\;.
\eeq

We also note that, after integration over the lepton angular variables $\bomega$,
we consistently obtain the Born level total cross section
${\hat \sigma}^{(0)}_{a_1 a_2 \to l_3l_4}(M^2)$ of the resummation formula
(\ref{resumydiff}). Indeed, after the $\bomega$ integration of 
Eq.~(\ref{distall}), the result does no longer depend on the lepton momenta and,
since it is a Lorentz invariant quantity, the result can only depends on the
invariant $(k_1+k_2)^2=2 k_1\cdot k_2=q^2=M^2$, which is independent of 
$\bkonet$. In other words, the dependence on the arbitrary parameter 
$\bkonet$ completely cancels in lepton-inclusive observables.

Using our $q_T$-recoil procedure, we can compute the corresponding lepton angular
function $F_{q_f {\bar q}_{f^\prime} \to l_3l_4}$ of Eq.~(\ref{Frecoil}).
This function is the product of two factors. One factor is a purely kinematical
origin (it derives from the phase space factor in the right-hand side of
Eq.~(\ref{distall})), and it depends on the specification of the angular variables
$\bomega$. The other factor, denoted as $F^{(D)}$ in the following (for simplicity
we omit the subscript $q_f {\bar q}_{f^\prime} \to l_3l_4$), has a dynamical
origin
and it depends on the Born level factor $| M^{(0)}(k_1,k_2;p_3,p_4)|^2$ in the 
right-hand side of Eq.~(\ref{distall}). Since $| M^{(0)}(k_1,k_2;p_3,p_4)|^2$
is a Lorentz invariant scalar quantity and the momenta $\{k_1,k_2;p_3,p_4\}$
are constrained by momentum conservation, 
$F^{(D)}$ can only depend on the dimensionless variable 
$4 k_1\cdot p_3/(2 k_1 \cdot k_2)=4 k_1\cdot p_3/M^2$.
Considering the centre--of--mass frame of $k_1$ and $k_2$, we have
$4 k_1\cdot p_3/M^2= 1- \cos \theta^\prime_{13}$, where $\theta^\prime_{13}$
is the scattering angle between $k_1$ and $p_3$. In other words, 
$F^{(D)}= F^{(D)}(\theta^\prime_{13})$ and $\theta^\prime_{13}$ is the lepton
scattering angle in a particular rest frame of the vector boson momentum $q^\mu$
(the centre--of--mass frame of $k_1$ and $k_2$). Our $q_T$-recoil procedure 
can thus be reinterpreted in terms of generation of lepton-pair events.
Considering a definite (with respect to the hadronic collision frame)
rest frame of the vector boson momentum, the lepton-pair event and the individual
lepton momenta are generated in that frame according to the corresponding Born
level angular distribution; then the lepton pair distribution is boosted to the
hadronic collision frame through the corresponding Lorentz transformation.
Since there is an infinite numbers of vector boson rest frames, this
event-generation procedure has an infinite degree of arbitrariness.
Applying a three-dimensional rotation to a vector boson rest frame, we obtain 
another vector boson rest frame and, thus, the infinite numbers of vector boson 
rest frames depends on the two scalar parameters of the 
three-dimensional rotation. Accordingly, our $q_T$-recoil procedure depends on the
arbitrary two-dimensional vector $\bkonet$, namely on two parameters
(the magnitude $|\bkonet|$ and the azimuthal angle of $\bkonet$).
In other words, the relation between the LO momenta $k_{i \,({\rm LO})}^\mu$
in Eq.~(\ref{lok12}) and the $q_T$-recoiled momenta $k_i$ obtained through
Eqs.~(\ref{k1res})--(\ref{ktmax}) can be reinterpreted as a Lorentz transformation
of the colliding parton momenta from the hadronic collision frame to a specified
vector boson rest frame. This interpretation directly relates our $q_T$-recoil
procedure with the specific CS frame procedure (as already mentioned and 
described in the initial part of this Appendix) that is directly used in other
resummed calculations \cite{Balazs:1995nz,Balazs:1997xd,Ellis:1997sc}. 
It can be explicitly checked that the 
CS frame procedure used to specify the lepton angular distribution 
$d{\hat \sigma}^{(0)}/d{\bomega}$ in the resummed calculation of 
Refs.~\cite{Balazs:1995nz,Balazs:1997xd,Ellis:1997sc} corresponds to the choice 
$\bkonet= \bktwot= \bqt/2$ (see Eqs.~(\ref{csk1}) and (\ref{csk2}))
within our class of $q_T$-recoil prescriptions. 


Owing to our explicit parametrization and implementation of the $q_T$-recoil
procedure, the quantitative effects of various $q_T$-recoil prescriptions
can be directly investigated in applications of the numerical program 
{\tt DYRes} (and of other resummed calculations). 
Obviously (as discussed in Sect.~\ref{sec:theory}),
different $q_T$-recoil prescriptions have no effects on quantities that are
fully inclusive over the leptonic variables $\bomega$. In general, our
expectations are as follows. We expect that the quantitative differences produced
by various $q_T$-recoil prescriptions are small for lepton non-inclusive
observables that are mostly sensitive to either the small-$q_T$ region
(in this region the $q_T$-recoil effects are non-singular and thus subdominant
with respect to the singular logarithmic contributions) 
or the high-$q_T$ region (in this region the $q_T$-recoil effects are suppressed
by the smooth switching procedure of Eqs.~(\ref{switch})--(\ref{fswitch})),
while relatively larger differences can appear in case
of sensitivity to the region of intermediate values of $q_T$.
Moreover, these quantitative differences are expected to decrease in going from 
NLL+NLO to NNLL+NNLO accuracy, since the non-vanishing $q_T$-recoil effects
start to formally contribute at the N$^{k+1}$LO level in the 
N$^{k}$LL+N$^{k}$LO calculation.
In Sect.~\ref{sec:lepton}
we have presented our quantitative results obtained with the 
{\tt DYRes} code. As stated at the beginning of Sect.~\ref{sec:results},
these results are obtained
(analogously to those in Refs.~\cite{Balazs:1995nz,Balazs:1997xd,Ellis:1997sc})
by computing the Born level angular distribution 
$d{\hat \sigma}^{(0)}/d{\bomega}$ in the CS rest frame, i.e. by setting 
$\bkonet= \bktwot= \bqt/2$ in the actual implementation of our
$q_T$-recoil procedure (this corresponds to use the prescription A in
Eqs.~(\ref{csz1})--(\ref{csk2})).
Setting $\bkonet \neq \bqt/2$, we have also
considered other variants of the $q_T$-recoil prescriptions 
and we have examined
the quantitative differences that are produced on the observables that are
examined in Sect.~\ref{sec:lepton} (i.e., the observables in 
Figs.~\ref{fig:Zqt}--\ref{fig:Wptlept}).
We have found that various $q_T$-recoil prescriptions produce differences that
are in agreement with our general expectations and, in particular, 
at NNLL+NNLO
accuracy these differences lead to small
quantitative effects: typically, the effects are much smaller than the
scale-variation uncertainties (estimated as in 
Sect.~\ref{sec:lepton}). For instance, comparing the $q_T$-recoil prescriptions
A (see Eqs.~(\ref{csz1})--(\ref{csk2})) and B (see Eq.~(\ref{k12asym})), we
obtain quantitative differences that are at most at the percent level
(e.g., in the case of the $\phi^*$ distribution of Fig.~\ref{fig:Wphi}
in the region $0.3 \ltap \phi^* \ltap 1$, and in the case of the lepton-$p_T$
and missing-$p_T$ distributions of Fig.~\ref{fig:Wptlept} in the region 
$45~{\rm GeV} \ltap p_T\ltap 50~{\rm GeV}$): these differences are definitely
smaller than the scale uncertainty (at both the NLL+NLO and NNLL+NNLO levels)
and, especially, they are of the same size as (and, hence, hardly distinguishable
from) the pure numerical errors of the {\tt DYRes} calculation in the 
NNLL+NNLO mode.

The $q_T$-recoil issue that we have discussed in this Appendix is not specific
of vector boson production and the ensuing leptonic decay. The issue affects
$q_T$ resummed calculations for any process
of the type
$h_1+h_2 \to {\cal F}(p_3,p_4,p_5,\dots)+X$ (we use the same notation as in
Eq.~(\ref{first})) where the final-state high-mass system ${\cal F}$ has total
transverse momentum $q_T$ and the momenta $p_3,p_4,p_5,\dots$ of its 
`decay products' are directly measured. Owing to the universality
(process-independent) structure of transverse-momentum resummation
\cite{Catani:2013tia}, the $q_T$-recoil procedure that we have introduced in this
Appendix  is directly applicable to all these processes.
The only key difference with respect to vector boson production is that the Born
level scattering amplitude  
$M^{(0)}(k_1,k_2;p_3,p_4)$ in Eq.~(\ref{distall}) is replaced by a properly
computable (all-loop) hard-virtual amplitude 
${\widetilde M}(k_1,k_2;p_3,p_4,p_5,\dots;\as(M^2))$ 
(see Sect.~4 in Ref.~\cite{Catani:2013tia}), which embodies QCD virtual radiative
corrections
(${\widetilde M}$ is computable as power series in $\as(M^2)$).
{\em Strictly speaking} \cite{Catani:2013tia}, the $q_T$ resummed cross section
at small values of $q_T$ is proportional to the angular dependent distribution
of the momenta $\{p_3,p_4,p_5,\dots \}$ as computed from ${\widetilde M}$
at $q_T=0$ (i.e., with the LO momenta $k_{i \,({\rm LO})}^\mu $ of 
Eq.~(\ref{lok12})). The ensuing $q_T$-recoil issue can be directly solved by our
$q_T$-recoil procedure. Indeed, the hard-virtual amplitude 
${\widetilde M}(k_1,k_2;p_3,p_4,p_5,\dots;\as(M^2))$ has the same kinematical
properties as its Born level counterpart ${\widetilde M}^{(0)}=M^{(0)}$:
therefore, the $q_T$ recoil can be directly implemented by simply evaluating 
${\widetilde M}(k_1,k_2;p_3,p_4,p_5,\dots;\as(M^2))$ with the $q_T$-recoiled
momenta $k_i$ of Eqs.~(\ref{k1res})--(\ref{ktmax}).

We add some final comments on spin correlations and on the specific process of SM
Higgs boson production and its decay in colourless particles
(e.g., $H \to \gamma \gamma, H \to WW \to \ell \nu\ell \nu, H \to ZZ \to 4\ell$)
\cite{deFlorian:2012mx}. The $q_T$ resummed Higgs boson cross section
at fixed momenta of the decay products is proportional to the angular distribution
as obtained (analogously to Eq.~(\ref{distall})) by the corresponding Born level
scattering amplitude 
$M^{(0)}_{g_1g_2 \to l_3 l_4 l_5 \dots}(k_1,k_2;p_3,p_4,p_5,\dots)$ 
for the gluon fusion process $gg \to H \to l_3 l_4 l_5 \dots$.
Owing to the spin-$0$ nature of the SM Higgs boson, $M^{(0)}$ factorizes in two
independent factors, 
$M^{(0)}_{g_1g_2 \to H}(k_1,k_2;q)$ and
$M^{(0)}_{H \to l_3 l_4 l_5 \dots}(q;p_3,p_4,p_5,\dots)$, for the production 
($g_1g_2 \to H$) and decay ($H \to l_3 l_4 l_5 \dots$) subprocesses of the 
Higgs boson. We thus have
$|M^{(0)}|^2= |M^{(0)}_{g_1g_2 \to H}(k_1,k_2;q)|^2 \,
|M^{(0)}_{H \to l_3 l_4 l_5 \dots}(q;p_3,p_4,p_5,\dots)|^2$. Note that 
$M^{(0)}_{H \to l_3 l_4 l_5 \dots}$ only depends on observable momenta, while 
$|M^{(0)}_{g_1g_2 \to H}(k_1,k_2;q)|^2$ only depends on $(k_1+k_2)^2=q^2$
because of Lorentz invariance. As a consequence, our  $q_T$-recoil procedure
(and its dependence on the definition of $k_1$ and $k_2$) has no effect on the
angular distribution of the Higgs boson decay products. The angular distribution
of the resummed calculation can be directly obtained \cite{deFlorian:2012mx}
by supplementing the Born level total cross section 
${\hat \sigma}^{(0)}_{gg \to H}(M^2)$ with the (kinematical and dynamical)
Higgs boson decay factor.
This specific example also clearly illustrates that the $q_T$-recoil issue that 
we have introduced in Sect.~\ref{sec:theory} and discussed in this Appendix is
directly related and due to the vector boson spin and the spin correlations
between the production and decay subprocesses of the vector boson. This kind of
relation between $q_T$ recoil and spin correlations is valid for generic
production processes.

\end{document}